\documentclass[a4paper,12pt]{article}

\pdfoutput=1
\usepackage{jheppub_X} 
\usepackage{hyperref}
\usepackage{lineno}
\usepackage{xcolor}
\usepackage{setspace}
\usepackage{subcaption}
\usepackage{float}
\usepackage{mathtools}
\usepackage{calc}

\usepackage[normalem]{ulem}

\usepackage{slashed}
\usepackage{array,booktabs}
\usepackage{multirow}

\usepackage{nicefrac}
\usepackage{dsfont}

\definecolor{darkorange}{rgb}{0.9,0.5,0.0}

\usepackage{cleveref}
\crefname{table}{Table}{Tables}
\crefname{equation}{Eq.}{Eqs.}
\crefname{appendix}{App.}{Apps.}
\crefname{section}{Sec.}{Secs.}
\crefname{figure}{Fig.}{Figs.}

\newcommand{\piDark}{\hat{\pi}}
\newcommand{\etaDark}{\hat{\eta}}
\newcommand{\rhoDark}{\hat{\rho}}

\title{Stopping Dark Mesons in Their Tracks with Long-Lived Particle and Resonant Signatures}

\affiliation[a]{Institute for Fundamental Science and Department of Physics, \\
University of Oregon, Eugene, OR 97403, USA}
\affiliation[b]{Department of Physics, University of California, San Diego, La Jolla, CA 92093, USA}
\affiliation[c]{Perimeter Institute for Theoretical Physics, 31 Caroline St N, Waterloo, ON N2L 2Y5, Canada}
\affiliation[d]{Laboratory for Elementary Particle Physics, Cornell University, Ithaca, NY 14853, USA}

\author[a]{Pouya~Asadi,}
\author[a]{Austin~Batz,}
\author[b]{Elias~Bernreuther,}
\author[c]{Marco~Costa,}
\author[d]{Samuel~Homiller,}
\author[a]{and Graham~D.~Kribs\hspace{0.8pt}}

\emailAdd{pasadi@uoregon.edu}
\emailAdd{abatz@uoregon.edu}
\emailAdd{ebernreuther@ucsd.edu}
\emailAdd{mcosta1@perimeterinstitute.ca}
\emailAdd{shomiller@cornell.edu}
\emailAdd{kribs@uoregon.edu}

\abstract{
Dark sectors with confining gauge interactions can provide both simple dark matter candidates and striking signals at colliders. We recast Large Hadron Collider searches for two different signatures of dark mesons that arise from a strongly-coupled theory with vector-like dark quarks that are in some non-trivial representation of Standard Model SU(2)$_L$. For any such electroweak representation, there is a 3-plet of dark mesons whose charged components are long-lived, and we reinterpret searches for disappearing tracks to place a lower bound on their mass of $\sim\!1.2\,\text{TeV}$. When the dark quarks are in SU(2)$_L$ representations larger than the fundamental, there is also a 5-plet of dark mesons that interacts with the electroweak gauge bosons via a chiral anomaly. We show that the 5-plet is the unique non-trivial meson multiplet with this anomaly and recast searches for the resulting diboson resonances to place bounds on model parameters. With additional measurements, the anomaly also enables one to reconstruct some ultraviolet parameters (the numbers of dark flavors and colors) while only measuring states in the infrared. Each of these signals represents an exciting opportunity for future searches using higher luminosity.
}

\begin{document}
\maketitle
\flushbottom

\section{Introduction}
\label{sec:intro}       

While new physics is required to explain dark matter,
the Standard Model (SM) has maintained remarkable consistency with various kinds of experiments over many years. 
Many searches for new physics have focused on minimal extensions of the SM and have placed stringent bounds on them.
This motivates exploring directions beyond the most minimal models. At the same time, simple models with concrete predictions serve as useful benchmarks to target in experimental searches. Confining dark sectors can simultaneously provide this non-minimality and simplicity. In such models, the SM is extended by a non-abelian gauge interaction with dark quarks in the ultraviolet (UV) that bind into dark hadrons in the infrared (IR). The UV theory may contain only a few degrees of freedom while the IR theory consists of a rich spectrum of states with a variety of possible interactions with the SM. 

New confining sectors have motivated numerous dedicated searches at the Large Hadron Collider (LHC) \cite{CMS:2021dzg,ATLAS:2023swa,CMS:2024gxp,ATLAS:2024xbu,CMS:2024nca,CMS:2025azd,CMS:2025mym,CMS:2025cwf,ATLAS:2025kuz,ATLAS:2025bsz}, as well as reinterpretations of other searches \cite{Barducci:2018yer,Kribs:2018ilo,Contino:2020god,Mies:2020mzw,Butterworth:2021jto}. 
Depending on the dark quark masses, confinement scale, number of flavors, number of colors, and the portal to the SM, these models can give rise to various intriguing signatures. Examples include long-lived particles \cite{Kilic:2009mi,Kilic:2010et,Harnik:2011mv,Schwaller:2015gea,Mahbubani:2017gjh,Buchmueller:2017uqu,Renner:2018fhh,Cheng:2019yai,Batz:2023zef}, 
novel shower patterns \cite{Strassler:2006im,Han:2007ae,Cohen:2015toa,Antipin:2015xia,Knapen:2016hky,Cohen:2017pzm,Mitridate:2017oky,Pierce:2017taw,Beauchesne:2017yhh,Bernreuther:2020vhm,Knapen:2021eip,Barron:2021btf,Cheng:2021kjg,Born:2023vll}, 
and exotic phenomena such as quirks \cite{Kang:2008ea,Knapen:2017kly,Evans:2018jmd}. 
See Refs.~\cite{Kribs:2016cew} and \cite {Albouy:2022cin} for reviews. Of particular importance to this work is the variety of potential dark hadron production mechanisms. 
New particles with electroweak interactions can of course be pair-produced at sufficiently high energies via Drell-Yan (DY) and vector boson fusion (VBF), but other possibilities including vector mesons mixing with gauge bosons \cite{Kilic:2009mi,Kribs:2018ilo} and single production of pion-like states through chiral anomalies \cite{Antipin:2015xia} exist as well. 
As we will show, correlations between the rates of these processes can even shed light on details of the UV theory.

We study collider signals of models with a confining dark sector that include vector-like dark quarks in the $N_f$-dimensional representation of SM SU(2)$_L$, in addition to being charged under the new SU($N_c$). 
See Refs.~\cite{Kilic:2009mi,Bai:2010qg,Antipin:2014qva,Appelquist:2015yfa,Mitridate:2017oky,Barducci:2018yer,Kribs:2018oad,Kribs:2018ilo,Abe:2024mwa} for similar models. 
It was recently discovered \cite{Asadi:2024bbq} that this model has a natural $\mathbb{Z}_2$ symmetry, known as $\mathcal{H}$-parity, that forbids the leading electromagnetic moments of the neutral dark hadrons. This observation on its own significantly relaxes na\"{i}vely stringent direct detection constraints on the dark baryons as dark matter candidates. It was additionally pointed out in Ref.~\cite{Asadi:2024tpu} that the lightest dark baryons are pure or approximate SM singlets for many combinations of $N_c$ and $N_f$, which even further suppresses direct detection signals of these so-called Noble Dark Matter candidates. 
The counter-intuitive implication is that while the dark quarks are electrically charged, the dark sector may be more readily discovered at colliders than other types of experiment. This is the possibility we investigate in this paper. Even beyond the underlying dark matter motivations, we will explore various interesting features of collider phenomenology that arise in the presence of heavy dark hadrons whose constituents are in an arbitrary SU(2)$_L$ representation.

Confining dark sectors with dark quarks that transform under the electroweak group exhibit rich spectroscopy, novel selection rules, and unconventional production and decay patterns that sharply contrast with more familiar non-confining dark sectors as well as sectors with SM-singlet dark quarks. Before turning to collider observables, we therefore delineate the key structural features of our model and identify the aspects most consequential for experimental searches. This discussion will set the stage for our main results: a systematic study of the collider signatures that arise once the dark sector becomes confining at scales relevant to the LHC.

We consider the regime where the mass of the dark quarks is significantly \emph{below} the dark sector confinement scale. This allows us to exploit our knowledge of SM quantum chromodynamics (QCD) and work in an  effective field theory (EFT) of dark mesons. We focus on the limit where the pseudo-Nambu--Goldstone bosons (pNGBs) of chiral symmetry breaking in the dark sector are the only dark hadrons that are kinematically accessible in experiments at the LHC. 

We analyze two under-explored signals of these confining dark sectors. The first is disappearing tracks from long-lived dark mesons, and the second is resonant diboson production from singly produced dark meson decays. These signatures allow us to reinterpret existing LHC searches to constrain our model parameters. Moreover, the model is an excellent benchmark for future searches targeting the same final states.

For all $N_f\geq2$, there is an SU(2)$_L$ 3-plet of pNGBs denoted by $\piDark_3$. For dark hadron mass scales within the regime we consider (\emph{i.e.}~above the scale of electroweak symmetry breaking), there is a $\sim\!170\,\text{MeV}$ mass splitting between the charged and neutral components of the $\piDark_3$ \cite{Cirelli:2005uq}. 
These mesons are forbidden from decaying directly to the SM by a $\mathbb{Z}_2$ known as $G$-parity (which is related to $\mathcal{H}$-parity and an analogous $\mathbb{Z}_2$ in the SM, as explained in \cref{sec:Gparity}) \cite{Bai:2010qg}. The dark $G$-parity has such importance that we denote $G$-odd pNGBs by $\piDark$ and $G$-even pNGBs by $\etaDark$, loosely following an analogy to SM mesons. 
The charged components of the $\piDark_3$ (denoted $\piDark_3^\pm$) decay to the neutral component $\piDark_3^0$ by emitting an SM pion, and the small phase space for this process causes the $\piDark_3^\pm$ to be long-lived. 
The $\piDark_3^\pm$ are therefore targets for long-lived particle (LLP) searches with signals very similar to those of charginos \cite{Barr:2002ex,Bomark:2013nya}. When the charged meson decays inside a detector's inner tracker and the soft pion is not reconstructed, this leaves a signature known as a disappearing track. In \cref{sec:tracks}, we use disappearing track searches to derive a lower bound on this meson's mass. 

We also study the single production of specific meson species.
If $N_f\geq3$, there is an SU(2)$_L$ 5-plet of pNGBs (denoted by $\etaDark_5$) in the spectrum that is \emph{unique} in that it is the only representation of dark mesons that has an anomaly with SU(2)$_L$. As explained in \cref{sec:anomaly}, this is analogous to the SM neutral pion's anomaly with quantum electrodynamics (QED).
This anomaly allows the $\etaDark_5$ to be resonantly produced via VBF before decaying to a pair of electroweak bosons. 
The rate for this process depends on $N_f$ and $f_\pi/N_c$, where $f_\pi$ is the dark pion decay constant. 
Thus, measuring these anomaly-induced collider signals would reveal details of the UV model. 
Similar examples of bound state resonances have been considered in Refs.~\cite{Nakai:2015ptz,Bai:2015nbs,Elor:2018xku,Bottaro:2021srh}. We use searches for scalar resonances decaying to diboson final states to place constraints on $f_\pi/N_c$ in \cref{sec:resonances}. 

\section{The Model}
\label{sec:model}

In this section, we outline the essential features of confining dark-sector models whose constituent dark quarks also transform under the SM SU(2)$_L$. Our aim is to underscore how seemingly innocuous aspects of the UV construction give rise to distinctive and unexpectedly rich collider phenomenology in the IR. While many of these ingredients have appeared previously in the literature, presenting them here in a unified framework provides the necessary basis for our comprehensive analysis of the novel collider signatures in the sections that follow.

We study dark mesons
arising from a new strongly-coupled confining gauge sector with the Lagrangian

\begin{equation}
\label{eq:L}
  \mathcal{L}_{\text{dark}} = -\frac{1}{4} G^{\mu\nu\, a}G_{\mu\nu}^a
  + \overline{\mathbf{Q}} \left(i\slashed{D}
  - m_0 \right)\!\mathbf{Q}\,,
\end{equation}

\noindent where $G_{\mu\nu}^{a}$ is the field strength of the dark SU($N_c$), and $\mathbf{Q}$ is a vector-like dark quark with mass $m_0$ that transforms as a fundamental of SU($N_c$) and an $N_f$-plet of SU(2)$_L$ with $N_f\geq2$. Since the quarks transform in the $N_f$-dimensional representation of SU(2)$_L$, one can think of SU(2)$_L$ as a gauged subgroup of an SU($N_f$) flavor symmetry. Gauge invariance requires the flavor multiplet to have degenerate mass at tree level. The quarks have vanishing hypercharge, so each dark sector particle has an electric charge equal to its weak isospin. At low energies, these quarks bind into dark mesons and baryons.

As pointed out in Ref.~\cite{Asadi:2024bbq}, this model has a $\mathbb{Z}_2$ symmetry known as $\mathcal{H}$-parity, under which the fields transform as

\begin{align}
    \mathbf{Q} &\stackrel{\vphantom{\int}\mathcal{H}}{\to} e^{i\pi J_y}\mathbf{Q}, & W_\mu^i &\stackrel{\vphantom{\int}\mathcal{H}}{\to} \mathcal{C}W_\mu^i\mathcal{C}, & B_\mu &\stackrel{\vphantom{\int}\mathcal{H}}{\to} \mathcal{C}B_\mu\mathcal{C}, \label{eq:H}
\end{align}

\noindent where $J_y$ is the second generator of SU(2)$_L$, $W_\mu^i$ is the mediator of SU(2)$_L$, $B_\mu$ is the mediator of U(1)$_Y$, and $\mathcal{C}$ denotes charge conjugation. 
The neutral dark quarks and hadrons in odd representations of SU(2)$_L$ are eigenstates of $\mathcal{H}$ with eigenvalue $\pm1$, whereas $\mathcal{H}$ maps charged states to opposite-charge states (up to a sign). $\mathcal{H}$-parity has great importance for the dark matter phenomenology of the dark baryons, as explored in Refs.~\cite{Asadi:2024bbq,Asadi:2024tpu}. 

This symmetry suppresses the dipole moments and charge radius of the dark baryons, as shown in~\cite{Asadi:2024bbq}. Even if dark matter is constituted by this stable bound state, direct detection searches would not be effective as a consequence of this suppression.
This observation motivates collider studies of this particular model
independent of its
cosmological history.

The next sections are devoted to highlighting the meson properties that are relevant for their collider phenomenology, namely their lifetimes and production rates.
In \cref{sec:chiral}, we describe the theory and spectrum of the mesons, which are the lightest bound states of the model, and hence better suited to be studied at colliders than the heavier baryons. In \cref{sec:Gparity}  we discuss another $\mathbb{Z}_2$ symmetry, distinct from $\mathcal{H}$-parity, known as $G$-parity \cite{Bai:2010qg} whose selection rules dictate large hierarchies between the meson lifetimes. While $\mathcal{H}$-parity does not play a role in collider phenomenology, $G$-parity is critical to understand decay chains of the dark mesons. 
In \cref{sec:mass} we discuss the mass splittings between the various mesons, which are crucial to compute rates for disappearing track signals.
While all mesons can be pair-produced at colliders through electroweak processes due to their SU(2)$_L$ charge, in \cref{sec:anomaly} we discuss the anomalous production channel that exists only for a specific meson multiplet, and that leads to a very peculiar resonant signature. This is a unique prediction of the strongly coupled theory discussed in this work that is not present for elementary SU(2)$_L$ particles. In \cref{sec:summary} we summarize these results.

\subsection{Mesons in the Chiral Effective Theory}
\label{sec:chiral}

This work focuses on the light quark limit, where the dark quark mass is significantly smaller than the dark sector confinement scale. In this limit, 
we can use an effective theory of dark mesons---the chiral Lagrangian---and trade  the UV parameters of the quark mass and confinement scale for the IR parameters of the lightest dark meson mass $m_\pi$ and the dark pion decay constant $f_\pi$. 

Since our regime is similar to that of the light mesons in QCD, we can borrow much intuition from the SM\@. The U($N_f$)$_L\times$U($N_f$)$_R$ flavor symmetry in the massless limit of \cref{eq:L} is spontaneously broken during confinement by the quark condensate $\langle \overline{\mathbf{Q}}\, \mathbf{Q}\rangle$ to its diagonal SU($N_f$) subgroup, plus a U(1) associated with a dark baryon number. There are $N_f^2-1$ light pseudo-scalar pNGBs in addition to a heavy flavor-singlet $\etaDark^\prime$ that is the would-be pNBG of the anomalous axial U(1). Explicit breaking of U($N_f$)$_L\times$U($N_f$)$_R$ by the quark masses and gauging its SU(2)$_L$ subgroup gives mass to the pNGBs, as further explained in \cref{sec:mass}. 

Each meson species 
transforms in some representation of SU(2)$_L$ that determines its interactions with the SM. A meson flavor state is a superposition of states with a quark flavor and an anti-quark flavor. Both the quark and the anti-quark are in the self-conjugate representation $\mathbf{N_f}$ of SU(2)$_L$, so the mesons are organized into the irreducible representations in 
\begin{equation}
\mathbf{N_f} \otimes \mathbf{N_f} = \bigoplus_{m=1}^{N_f}(\mathbf{2m-1}) = \mathbf{1} \oplus \mathbf{3} \oplus \mathbf{5} \oplus \cdots \oplus (\mathbf{2N_f-1}). \label{eq:reps}
\end{equation}
\noindent For the lowest-lying pseudo-scalar mesons, we can identify the $\mathbf{3},\mathbf{5}$, \emph{etc.}~as the representations of the pNGBs and the flavor-singlet $\mathbf{1}$ as the would-be pNGB $\etaDark^\prime$. Note in particular that there is always a 3-plet $\piDark_3$ (since $N_f\geq2$), and there is a 5-plet $\etaDark_5$ for all $N_f\geq3$. 
In the rest of this work, we explore the interesting features that arise for each of these meson representations and the consequences for phenomenology. 

\subsection{Dark \texorpdfstring{$G$}{G}-Parity}
\label{sec:Gparity}

This model has a  $\mathbb{Z}_2$ symmetry known as $G$-parity \cite{Bai:2010qg} that has important  implications for the dark mesons and has an analogy in the SM \cite{Lee1956,PDG:2024cfk}. 
Let us first review the SM case. The isospin group SU(2)$_I$, which acts on $u$ and $d$ quarks, is a subgroup of the SU(3) flavor symmetry that acts on $u$, $d$, and $s$ quarks.\footnote{SU(2)$_I$ is embedded in flavor SU(3) in the SM differently from how SU(2)$_L$ is embedded in SU($N_f$) in the dark sector. In our model, the fundamental of SU($N_f$) is identified with the $\mathbf{N_f}$ of SU(2)$_L$, while the fundamental of flavor SU(3) in the SM is identified with $\mathbf{1} \oplus \mathbf{2}$ of SU(2)$_I$.} Besides the would-be pNGB SM $\eta^\prime$, there is an SU(3) octet of pNGBs containing an iso-triplet (the pions), two iso-doublets (the kaons), and an iso-singlet (the SM $\eta$).\footnote{More precisely, the SM $\eta$ and $\eta^\prime$ mass eigenstates are admixtures of the flavor singlet and flavor octet. We comment on mass mixings of dark sector flavor states in \cref{sec:mass}.} The SM $G$-parity transformation is $G_{\text{SM}}=\mathcal{C}e^{i\pi I_y}$ (where $I_y$ is the second generator of SU(2)$_I$), which has eigenstates 
\begin{align}
    \eta^\prime_{\,\text{SM}} &\stackrel{\vphantom{\int} G_{\text{SM}}}{\to} \eta^\prime_{\,\text{SM}}, & \pi_{\text{SM}} &\stackrel{\vphantom{\int} G_{\text{SM}}}{\to} -\pi_{\text{SM}}, & \eta_{\,\text{SM}} &\stackrel{\vphantom{\int} G_{\text{SM}}}{\to} \eta_{\,\text{SM}},
\end{align}
\noindent among the lowest-lying pseudo-scalar mesons. The $G_{\text{SM}}$ charge of a more general meson is $(-1)^{I+\ell+s}$, where $I\in\{0,1\}$ is the isospin, $\ell$ is the orbital angular momentum of the constituent quarks, and $s$ is the total spin of the constituent quarks \cite{PDG:2024cfk}. SM $G$-parity is preserved by QCD, which leads to selection rules such as the forbidding of the $\eta_{\,\text{SM}}\to3\pi_{\text{SM}}^0$ decay as a strong process \cite{Nefkens:2002sa}. However, SM $G$-parity is violated at tree-level by QED. After all, $\eta_{\,\text{SM}}\to3\pi_{\text{SM}}^0$ is one of the principal decay modes of the $\eta_{\,\text{SM}}$. By contrast, $G$-parity in the dark sector provides much stronger selection rules.

The dark $G$-parity transformation on dark sector matter content is \cite{Bai:2010qg}
\begin{equation}
    G = \mathcal{C}e^{i\pi J_y} = \mathcal{C}\mathcal{H},
\end{equation}
\noindent where the isospin rotation in the SM $G$-parity definition has been replaced by a global SU(2)$_L$ rotation.\footnote{$G$ also acts as charge conjugation on the dark gluon field so that the dark SU($N_c$) coupling is $G$-even \cite{Bai:2010qg}.} Clearly, $G$ and $\mathcal{H}$ are closely related. A key difference is that $G$ maps particles to anti-particles, while $\mathcal{H}$ maps a particle to a different particle in the same electroweak multiplet, up to a sign. In the case of baryons, $G$ maps a baryon to an anti-baryon, so a baryon cannot be $G$ eigenstate even if it can be an $\mathcal{H}$ eigenstate. Mesons, however, have particles and anti-particles \emph{in the same electroweak multiplet}, so they can be eigenstates of $G$, $\mathcal{H}$, or both.

Consider a meson state $|J, \, Q\rangle$ in the $2J+1$-dimensional representation of SU(2)$_L$ with electric charge $Q$. That is, 
\begin{align}
    J_z|J, \, Q\rangle &= Q|J, \, Q\rangle\, , \\
    J^2|J, \, Q\rangle &=J(J+1)|J, \, Q\rangle\,,
\end{align}
\noindent where $J_z$ is the third generator of SU(2)$_L$, and $J^2$ is the quadratic Casimir operator. Following the SM analogy, the actions of the various $\mathbb{Z}_2$ transformations can be self-consistently written as
\begin{align}
    \mathcal{H}|J, \, Q\rangle &= \mathmakebox[\widthof{$(-1)^{Q+s+\ell}$}][r]{(-1)^{J-Q}} |J,\,-Q \rangle \, ,\label{eq:Hrule} \\
    \mathcal{C}|J, \, Q\rangle &= \mathmakebox[\widthof{$(-1)^{Q+s+\ell}$}][r]{(-1)^{Q+s+\ell}}|J,\,-Q \rangle\, ,\label{eq:Crule} \\
    G |J, \, Q\rangle &= \mathmakebox[\widthof{$(-1)^{Q+s+\ell}$}][r]{(-1)^{J+s+\ell}} |J, \, Q\rangle\,. \label{eq:Grule}
\end{align}
\noindent One can see from these transformation rules that \emph{every meson state in a particular electroweak multiplet is an eigenstate of $G$-parity}. We focus on the pseudo-scalar pNGBs in this work, so we set $s=\ell=0$ from now on. 

Recall from \cref{eq:reps} that the mesons appear in the $\mathbf{1},\mathbf{3},\mathbf{5},\ldots$ representations of SU(2)$_L$. Therefore, \cref{eq:Grule} implies that the meson multiplets have alternating $G$-parity, \emph{i.e.}~the singlet, 5-plet, 9-plet, \emph{etc.}~are $G$-even, and the 3-plet, 7-plet, 11-plet, \emph{etc.}~are $G$-odd. 
To make our meson naming scheme explicit, we denote the $G$-odd pNGBs as $\piDark$ and the $G$-even pNGBs as $\etaDark$.
We also specify the SU(2)$_L$ representations of each meson with subscripts and electric charges with superscripts when necessary. 
The different types of mesons, their SU(2)$_L$ representations, the minimum $N_f$ for them to exist, and their behavior under dark $G$-parity are summarized in Table~\ref{tab:species}.
Again, we emphasize that only the neutral components of electroweak multiplets have well-defined $\mathcal{H}$-parity, while entire meson multiplets have well-defined $G$-parity.

\begin{table}[t]
\centering
\begin{tabular}{>{$}c<{$} | >{$}c<{$} | >{$}c<{$} | >{$}c<{$}}
\hline
\text{Species} & \text{SU(2)}_L ~ \text{Rep.} & N_{f,\text{min}} & G\text{-parity} \\
\hline
\multirow{1}{*}{$\piDark_3$} & \multirow{1}{*}{$\mathbf{3}$} & \multirow{1}{*}{2} & \multirow{1}{*}{$-1$}  \\
\hline
\etaDark_5 & \mathbf{5} & 3 & +1   \\
\hline
\piDark_7,\, \piDark_{11},\, \piDark_{15},\dots & \mathbf{7}, \mathbf{11}, \mathbf{15}, \dots & 4,6,8,\dots & -1  \\ 
\cline{1-4}
\etaDark_9,\, \etaDark_{13},\, \etaDark_{17},\dots & \mathbf{9}, \mathbf{13}, \mathbf{17}, \dots & 5,7,9,\dots & +1   \\ 
\hline
\end{tabular}
\caption{\label{tab:species} Summary of the pNGBs of dark chiral symmetry breaking, including their representation under SU(2)$_L$, the minimum $N_f$ such that the species exists in the spectrum (see \cref{sec:chiral}), and the charge under the dark $G$-parity (see \cref{sec:Gparity}).}
\end{table}

While strong dynamics preserve $G$-party, in general interactions with heavier degrees of freedom could break it. 
While we do not focus on the cosmology of the model, it is noteworthy to show that meson decays induced by these $G$-violating interactions (which are expected to exist) do not spoil the standard cosmological picture, and moreover do not significantly affect collider signatures.

Dark $G$-parity can be broken by the dimension-5 effective operators  \cite{Bai:2010qg}
\begin{align}
B_{\mu\nu} \overline{\mathbf{Q}}\, \sigma^{\mu\nu}\,\mathbf{Q} & \; \stackrel{\vphantom{\int}G}{\to} \; B_{\mu\nu} \left( \mathcal{C}\overline{\mathbf{Q}} \,\sigma^{\mu\nu}\,\mathbf{Q}\mathcal{C} \right) \; = \; - B_{\mu\nu} \overline{\mathbf{Q}}\, \sigma^{\mu\nu}\,\mathbf{Q}\, , \\
     H^\dagger \tau^a H \, \overline{\mathbf{Q}}\, J^a \,\mathbf{Q} & \; \stackrel{\vphantom{\int}G}{\to} \; H^\dagger \tau^a H \, \overline{\mathbf{Q}} \left( e^{-i\pi J_y} J^a e^{i\pi J_y} \right)\mathbf{Q} \; = \; H^\dagger \tau^a H \, \overline{\mathbf{Q}} \left( -J^a\right)^\ast \mathbf{Q},
\end{align}
\noindent where $B_{\mu\nu}$ is the hypercharge field strength, $H$ is the Higgs doublet, and $\tau^a$ is a generator of the fundamental representation of SU(2)$_L$. However, unlike SM $G$-parity, there are no gauge-invariant renormalizable operators that violate the dark $G$-parity.\footnote{$\mathcal{H}$-parity is violated at two-loop order by the SM due to its action on the hypercharge field \cite{Asadi:2024bbq}. By contrast, $G$ leaves all SM fields invariant.} This quasi-stabilizes the lightest $G$-odd meson, which is the $\piDark_3^0$. 
As noted in Refs.~\cite{Bai:2010qg,Asadi:2024bbq}, if these operators are suppressed by the Planck scale, then the quasi-stable $\piDark$ mesons produced in the early universe may survive past Big Bang nucleosynthesis but be too short-lived to be cosmologically safe.
By dimensional analysis, we can roughly estimate the scaling of the $\piDark_3^0$ lifetime $\tau_3$ with the scale of $G$-parity breaking $\Lambda_{\slashed{G}}$ as 
\begin{align}
\label{eq:pi30_lifetime}
\tau_3 &\sim \frac{a^2\Lambda_{\slashed{G}}^2}{m_\pi^3} \, , \\
\Lambda_{\slashed{G}} &\sim \frac{10^{16}\,\text{GeV}}{a} \sqrt{\left(  \frac{m_\pi}{1\,\text{TeV}} \right)^3\left( \frac{\tau_3}{0.1\,\text{s}} \right)}\,,
\end{align}
\noindent where $a>1$ is a dimensionless coefficient encapsulating coupling and phase space factors. 
To ensure $\tau_3\lesssim0.1\,\text{s}$ (so that the decays do not disrupt nucleosynthesis~\cite{Jedamzik:2006xz,Kawasaki:2020qxm}), $\Lambda_{\slashed{G}}$ should be below $\sim \! 10^{16}\,\text{GeV}$ when $m_\pi \sim \mathcal{O}(1)$\,TeV. Therefore, we need some UV completion below the Planck scale to facilitate these decays. It was also noted in Ref.~\cite{Asadi:2024tpu} that a UV completion below the Planck scale is required for many values of $N_c$ and $N_f$ to address sub-Planckian Landau poles of the electroweak coupling, and such a UV completion could also provide the needed $G$ violation. While these $G$-parity violating decays may themselves have interesting phenomenology and help probe the scale of the UV completion, we do not consider them in this work. The lifetime could easily be well beyond the detector scale, and so we treat the $\piDark_3^0$ as stable in our collider study in \cref{sec:signals}. 

\subsection{Meson Masses}
\label{sec:mass}

The chiral effective theory we employ has a regime of validity up to a UV mass scale of $\sim\!4\pi f_\pi$. We therefore restrict our discussion to the range of pNGB masses below that scale, but there is some subtlety in the mass spectrum of the different pseudo-scalar species. The pNGBs get their masses from explicit breaking of chiral symmetry, of which there are two sources: quark masses and the gauging of the SU(2)$_L$ subgroup. The quark masses are degenerate at tree-level, so the pNGB masses are as well. Mass splittings arise due to loops with electroweak bosons of the form

\begin{figure}[H]
\centering
\begin{minipage}[t]{0.25\textwidth}
  \vspace{0em}
  \includegraphics[width=\linewidth]{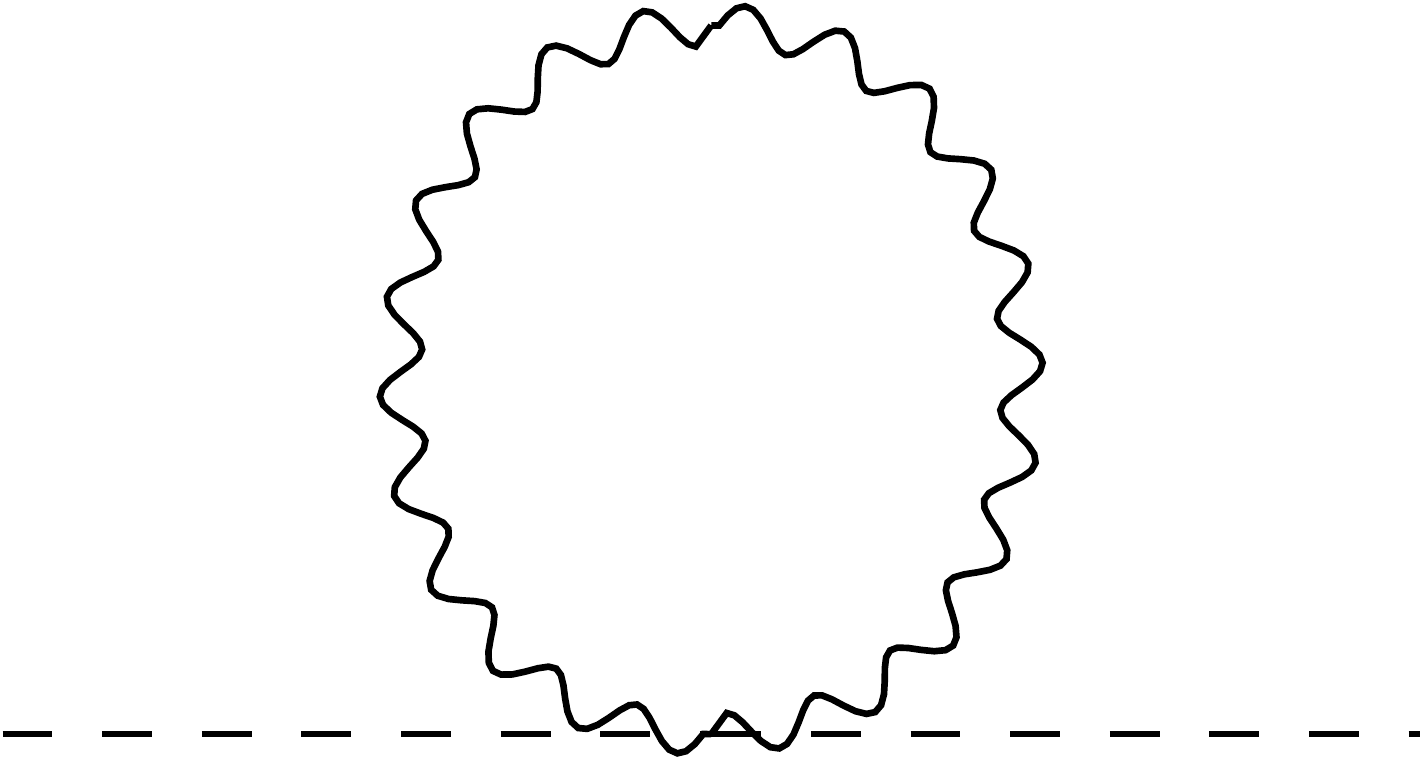}
\end{minipage}
\hspace{0.05\textwidth} 
\begin{minipage}[t]{0.25\textwidth}
  \vspace{2.52em}
  \includegraphics[width=\linewidth]{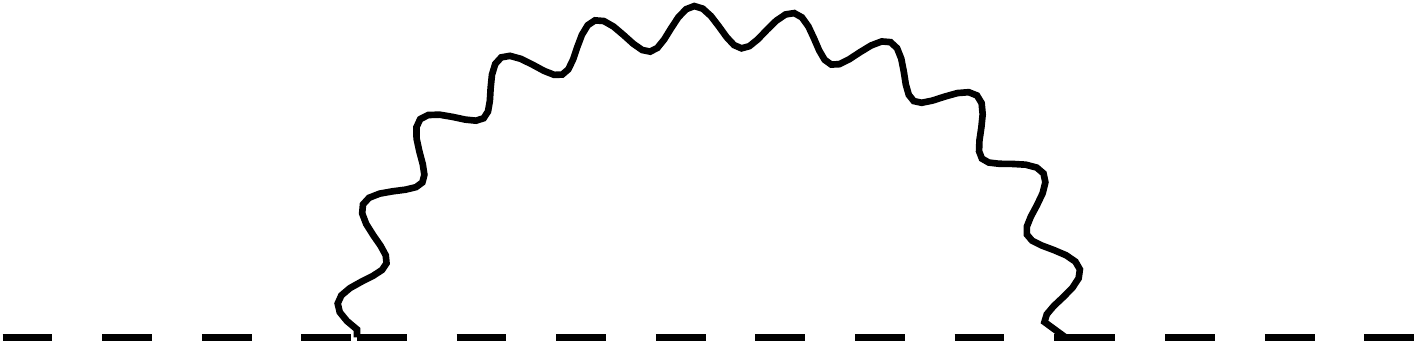}
\end{minipage}
\end{figure}

\noindent that depend on the meson charge and representation. First, consider the mass splitting between mesons in different electroweak multiplets. The diagram on the left is quadratically divergent, so a meson in the $2J+1$-dimensional representation of SU(2)$_L$ gets a dominant correction to its mass-squared of $\sim\!4\pi \alpha_W J(J+1)f_\pi^2$, where $\alpha_W$ is the SU(2)$_L$ coupling \cite{Bai:2010qg}. Even in the massless quark limit, this prevents $m_\pi$ from being much smaller than $f_\pi$ without fine-tuning. Then, the difference in the squared masses of the pNGB multiplets scales like the difference in their quadratic Casimirs, but we cannot precisely compute it in perturbation theory. 

Next, consider the mass splitting between mesons with different charges in the same electroweak multiplet. The same kinds of loops contribute, but the splitting is due to electroweak symmetry breaking rather than the cutoff of the EFT. The calculation is performed in Ref.~\cite{Cirelli:2005uq}, and the splitting is $\sim\!170\,\text{MeV}$ between singly-charged and neutral components of a multiplet for meson masses above $\mathcal{O}(100)\,\text{GeV}$.

The splittings between components of the electroweak multiplets have a strong impact on the $\piDark$ meson lifetimes, as they are forbidden from decaying directly to the SM by $G$-parity. Therefore, charged $\piDark$ mesons decay via cascade, \textit{i.e.}~they decay to a lighter species in the same electroweak multiplet by radiating one or two off-shell gauge bosons, and the lighter species (if charged) then decays in the same way. This implies the singly-charged $\piDark$ mesons can decay as $\piDark^\pm\to \piDark^0 \ \pi_{\text{SM}}^\pm$, as the far off-shell $W^\pm$ radiated by the $\piDark^\pm$ effectively mixes with the SM pion. There is additional discussion of this process in Refs.~\cite{Belyaev:2016lok,Belyaev:2020wok}, and the resulting decay rate is\footnote{As discussed in Ref.~\cite{Belyaev:2016lok}, decays with leptonic final states are suppressed to the extent that the charged state would be collider-stable if the hadronic channel were not open. This is in part due to multi-body phase space suppression.}

\begin{equation}
    \Gamma(\piDark^\pm\to\piDark^0\ \pi^\pm_{\text{SM}}) = \pi \,\alpha_W^2  \frac{f_{\pi,\text{SM}}^2}{m_W^4} J(J+1) \Delta m_{+,0}^2\sqrt{\Delta m_{+,0}^2-m_{\pi,\text{SM}}^2}\,, \label{eq:piRate}
\end{equation}

\noindent where $f_{\pi,\text{SM}}\simeq 92\,\text{MeV}$ is the SM pion decay constant, $J$ indicates the $\piDark$ meson representation, $m_W$ is the $W$ boson mass, $\Delta m_{+,0}$ is the splitting between the singly-charged and neutral $\piDark$ mesons, and $m_{\pi,\text{SM}}$ is the SM pion mass.\footnote{One can compute the corresponding rate for a higher-charge $\piDark$ meson to decay to $\piDark$ meson with one less charge by replacing $J(J+1)$ in \cref{eq:piRate} with  the square of the corresponding element of the SU(2)$_L$ ladder operator in the appropriate representation.} The small phase space for this decay causes the $\piDark_3^+$ to be long-lived on a collider scale with proper lifetime $\tau$ roughly between $4\,\text{cm}$ and $6\,\text{cm}$ (depending on the mass). We explore the consequences for phenomenology in \cref{sec:tracks}. 

The mass splittings between different multiplets implies that the higher multiplets can decay by ``hopping'' down to a lower one. These processes are permitted by loop diagrams of the form \cite{Bai:2010qg}

\begin{figure}[H]
\centering
\begin{minipage}[t]{0.28\textwidth}

  \includegraphics[width=\linewidth]{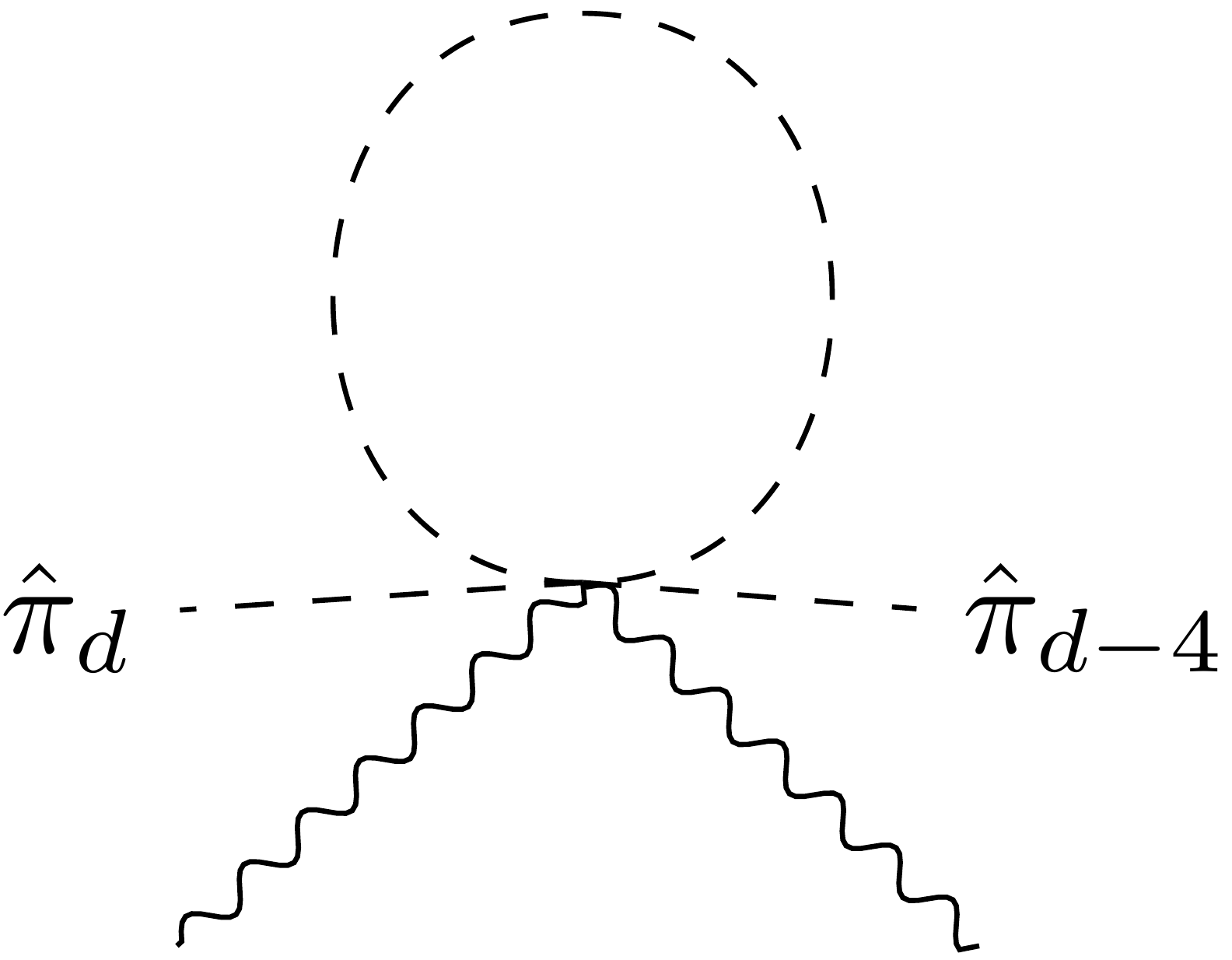}
\end{minipage}
\hspace{0.1\textwidth} 
\begin{minipage}[t]{0.28\textwidth}

  \includegraphics[width=\linewidth]{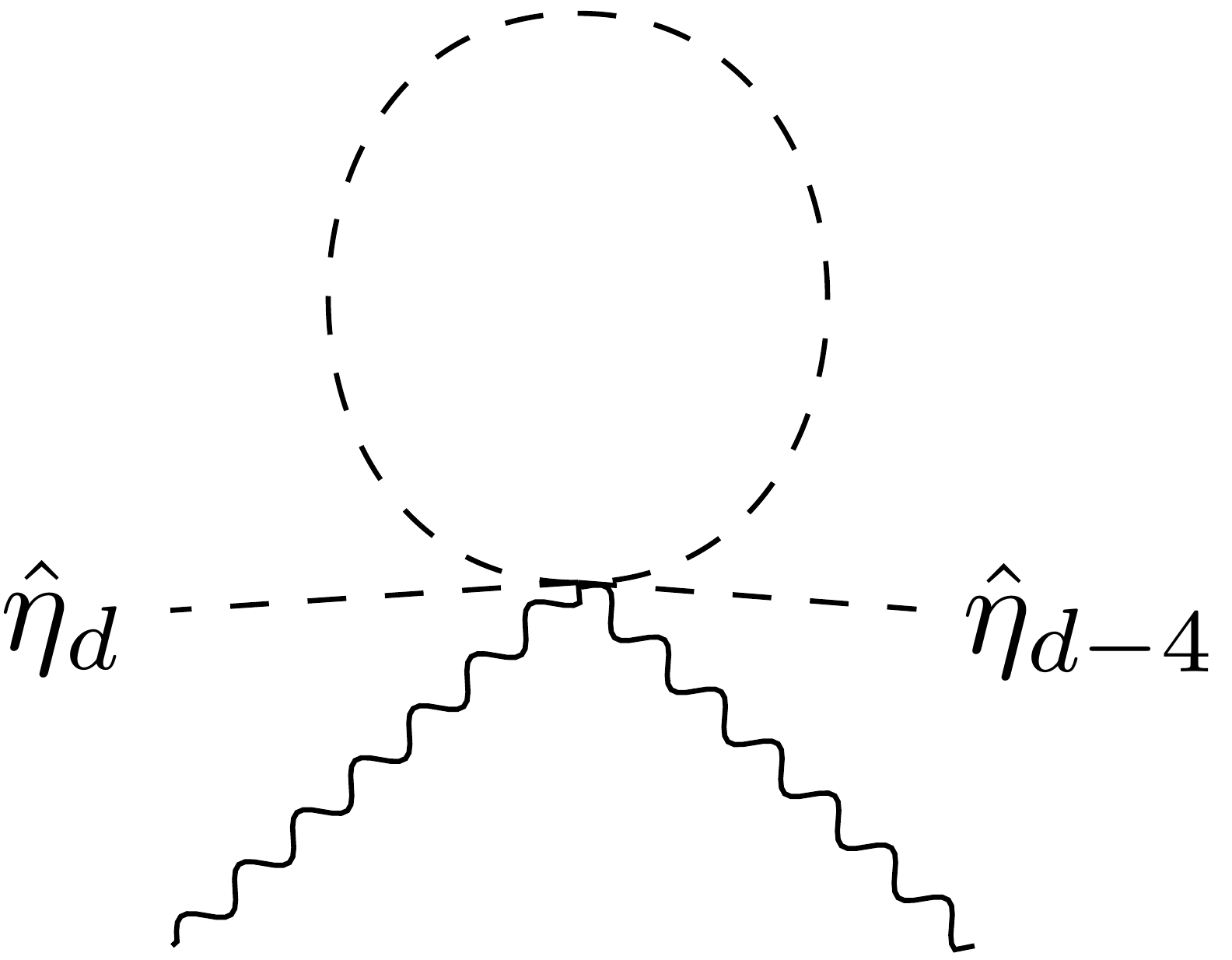}
\end{minipage}

\end{figure}

\noindent with some dark meson species in the loop. These six-point vertices arise from dimension-6 effective operators in the meson EFT. In this way, mesons in representations larger than the $\mathbf{5}$ can hop down to a meson in a lower representation with the same $G$-parity by emitting two electroweak bosons. Since these decays are facilitated by dimension-6 operators, we expect the decay rate to scale with the model parameters as

\begin{equation}
    \Gamma_{\text{hop}} \sim \alpha_W^2\frac{\Delta m_{d,d-4}^5}{(4\pi)^5 f_\pi^4}\,,
    \label{eq:hop}
\end{equation}

\noindent where $\Delta m_{d,d-4}$ is the splitting between the initial and final meson multiplets. While this decay rate is suppressed by a loop factor and multi-body phase space, the mass splitting itself is expected to scale as $f_\pi$. Moreover, the divergence of the loop introduces an additional numerical enhancement. Therefore, we expect the higher multiplets to rapidly decay to the $\piDark_3$ or $\etaDark_5$ in this way.

This work focuses on the limit where the pNGBs are light enough to be accessible at the LHC while the heavier mesons (the $\etaDark^\prime$, vector mesons $\rhoDark$, \emph{etc.}) are inaccessible. The reason is to avoid various complications such as the $\rhoDark$ mixing with the SM $Z$ boson \cite{Kribs:2018ilo}, interactions of the pNGBs with the $\rhoDark$, and the tower of other excited meson states. Indeed, many excited QCD resonances appear within an $\mathcal{O}(1)$ factor of the SM $\rho$ mass \cite{PDG:2024cfk}, so we expect the same to be true in the dark sector. We generically expect these heavier meson masses to be $\sim\!4\pi f_\pi$, so we will take $4\pi f_\pi$ to be above some threshold (below which some resonance searches may be sensitive) to safely ignore them. One possible exception, depending on the details of the UV, is the $\etaDark^\prime$, which becomes light in the large-$N_c$ limit \cite{WITTEN1980363}. A light $\etaDark^\prime$ would slightly alter phenomenology for reasons that will become clear in \cref{sec:anomaly}, but we neglect this possibility in the rest of this work. 

To get a rough estimate of the minimum mass where the $\rhoDark$ may be visible, we can compare our model to the similar dark sector described in Refs.~\cite{Kribs:2018oad} and \cite{Kribs:2018ilo} with dark vector mesons mixing with SM gauge bosons. There are dilepton resonance searches \cite{ATLAS:2017fih,CMS:2018ipm}, as well as the search in Ref.~\cite{ATLAS:2024xbu} for the vector meson in the similar model decaying to pairs of pNGB mesons. These searches probe the vector mass scale up to $\sim\!2\,\text{TeV}$ at most, so this serves as a rough lower bound on $4\pi f_\pi$ above which the signals discussed in this work are dominant.

Another possible complication is mixing of mesons in different electroweak multiplets. This is in some sense analogous to SM $\eta$-$\eta^\prime$ mixing, but it is clearly only permitted in the dark sector by electroweak symmetry breaking. Ref.~\cite{Asadi:2024tpu} shows how mixing of baryon multiplets is important to understanding their dark matter phenomenology, but the effect disappears as the relevant mass scales in the dark sector become larger than the electroweak scale. Also, if any of the dark sector mass scales were below the electroweak scale, then the details of the electroweak interactions between individual quarks may significantly alter the meson spectrum, as was the case for baryons in Ref.~\cite{Asadi:2024tpu}. We neglect these effects in this work, as the relevant scales are set by $4\pi f_\pi$.

In summary, the mass splitting between the multiplets is set by the Casimirs and $f_\pi$, the splitting between entries within a multiplet is small and leads to long-lived $\piDark_3^\pm$ states, and the pNGBs are the only mesons we consider in our collider study. We have yet to discuss the decays of the $\etaDark_5$ mesons, for which we need the findings of \cref{sec:anomaly}. This will show that each meson multiplet has its own idiosyncrasies with direct impact on experimental signals. 

\subsection{The 5-plet Anomaly}
\label{sec:anomaly}

This section presents the striking result that not only can the $\etaDark_5$ be singly produced and decay via an anomaly with SU(2)$_L$, but also that it is \emph{unique} among the pNGBs in this way. The gauge-invariant form of the Wess-Zumino-Witten \cite{Wess:1971yu,Witten:1983tw,Witten:1983ar} operator contains the dimension-5 effective operator \cite{Peskin:1995ev,Bai:2010qg,Tong2018}
\begin{equation} 
    \label{eq:Lanomaly}
    \mathcal{L}_{\text{anomaly}} = \frac{g_W^2}{16\pi^2} \frac{N_c}{f_\pi} \hat{\phi}^a \varepsilon^{\mu\nu\alpha\beta} W^i_{\mu\nu} W^j_{\alpha\beta} \text{Tr} \left[ T^a J^i J^j  \right],
\end{equation}
\noindent where $g_W=\sqrt{4\pi\alpha_W}$, $W^i_{\mu\nu}$ is the SU(2)$_L$ field strength, $J^i$ is a generator of SU(2)$_L$, the sum over the index $a$ is a sum over all pNGBs $\hat{\phi}$, and $T^a$ is the generator of flavor SU($N_f$) corresponding to the $a^{\text{th}}$ pNGB. It has been noticed that this operator vanishes for the $\piDark_3$ \cite{Bai:2010qg,Howe:2016mfq,Barducci:2018yer}, and Refs.~\cite{Bai:2010qg,Barducci:2018yer} remarked that the $G$-even mesons can decay in this way, but the fact that the $\etaDark_5$ is unique has not appeared in the literature. That is, if $N_f$ is large enough such that an $\etaDark_{9},\etaDark_{13}$, \emph{etc.}~exist in the spectrum, they 
do not have an anomaly with $SU(2)_L$, and so they do not decay like the $\etaDark_{5}$.
Recall from \cref{sec:chiral} that the $\etaDark_5$ only exists in the spectrum if the quarks are in larger representations of SU(2)$_L$ than the fundamental, so studies with constituents in exclusively fundamental representations of their gauge groups would have missed this mechanism.

In \cref{sec:SManomalies} we review SM chiral anomalies, while in \cref{sec:uniqueness} we present a detailed argument that the $\mathbf{5}$-plet is the \textit{unique} pNGB for which this operator is non-vanishing. To summarize the proof, the operator is forbidden for representations larger than the $\mathbf{5}$ by gauge invariance, and it vanishes for the $\mathbf{3}$ due to the vanishing of anomaly coefficients of SU(2) representations.

\begin{figure}
    \centering
    \includegraphics[width=0.28\textwidth]{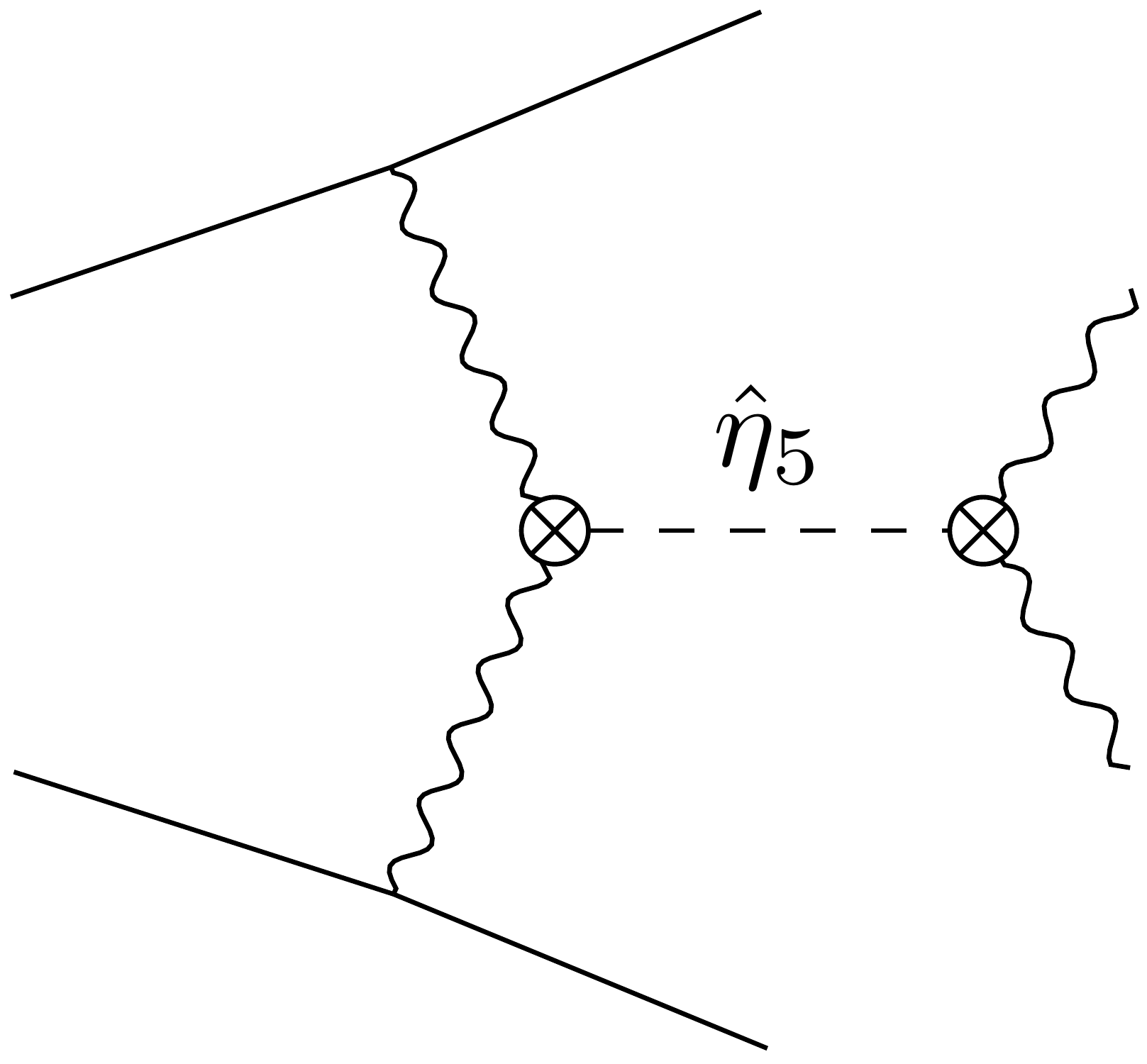}
    \caption{Feynman diagram corresponding to the anomaly-induced resonant production of the $\etaDark_5$. The $\otimes$ symbols denote  insertions of the anomaly operator in \cref{eq:5pletAnomaly}. The $\etaDark_5$ species are the unique pNGBs of dark chiral symmetry breaking with this operator. 
    Observation of this signal provides sensitivity to UV parameters such as $N_c$ and $N_f$, which set the strength of the coupling (see \cref{eq:5pletAnomaly,eq:c_value}). LHC implications are discussed in \cref{sec:resonances}. }
    \label{fig:anomaly}
\end{figure}

Then, we can expand \cref{eq:Lanomaly} as
\begin{align}
    \label{eq:5pletAnomaly}
    \mathcal{L}_{\text{anomaly}} = c \,\frac{g_W^2}{16\pi^2} \frac{N_c}{f_\pi} \varepsilon^{\mu\nu\alpha\beta} \left[ 
    \sqrt{\frac{3}{2}}\right.\, &\etaDark_5^{++}\,(W^1_{\alpha\beta}+iW^2_{\alpha\beta})(W^1_{\mu\nu}+iW^2_{\mu\nu})
      \\
     - \sqrt{6}\,&\etaDark_5^+\,W^3_{\mu\nu}(W^1_{\alpha\beta}+iW^2_{\alpha\beta}) \notag \\ 
     + \, &\etaDark_5^0\,(2W^3_{\mu\nu}W^3_{\alpha\beta} - W^1_{\mu\nu}W^1_{\alpha\beta} - W^2_{\mu\nu}W^2_{\alpha\beta}) \notag \\
    + \sqrt{6}\,&\etaDark_5^-\,W^3_{\mu\nu}(W^1_{\alpha\beta}-iW^2_{\alpha\beta}) \notag \\
    + \sqrt{\frac{3}{2}}\, &\etaDark_5^{--}\,(W^1_{\alpha\beta}-iW^2_{\alpha\beta})(W^1_{\mu\nu}-iW^2_{\mu\nu})\left.\vphantom{\sqrt{\frac{3}{2}}} \right], \notag
\end{align}
\noindent where the relative factors between the terms can be found using the techniques in \cref{sec:generators}, and
\begin{equation}
    c = \frac{1}{12\sqrt{10}}\sqrt{\frac{(N_f+2)!}{(N_f-3)!}}
    \label{eq:c_value}
\end{equation}
\noindent is an overall coefficient derived in \cref{sec:c}. Since $c$ scales as $N_f^{5/2}$, the strength of the anomaly interaction increases rather quickly with $N_f$. In the electroweak-broken phase, this operator leads to the decay of the $\etaDark_5$ to different combinations of $W^\pm$, $\gamma$, and $Z$ boson pairs.\footnote{Since $W_{\mu\nu}^i$ is the full SU(2)$_L$ non-abelian field strength, these operators also include couplings of the $\etaDark_5$ to three electroweak bosons at higher order in $g_W$.} 
In addition, the anomaly also allows the $\etaDark_5$ to be \emph{resonantly produced} in proton-proton collisions via VBF \cite{Antipin:2015xia},\footnote{See Ref.~\cite{Dugan:1984hq} for another example (in a composite Higgs context) of having a subset of mesons that can be produced through VBF via an anomaly.} followed by decay to pairs of electroweak bosons, as shown in \cref{fig:anomaly}. 
The collider phenomenology of this global-gauge-gauge anomalous interaction is discussed in \cref{sec:resonances}.

The anomaly is the dominant decay mode of each species in the 5-plet. One may have expected the singly-charged $\etaDark^\pm$ mesons to decay analogously to the SM pion, \textit{i.e.}~via effectively mixing with the $W$ boson. However, the operator $W^\pm_\mu \partial^\mu \pi^\mp_{\text{SM}}$ that leads to this mixing is parity-violating. The couplings of the $W$ to the SM quarks violate parity, so this mixing operator naturally appears in the chiral Lagrangian. By contrast, the dark quarks have vector-like couplings to the $W$, so operators such as $W^\pm_\mu \partial^\mu \etaDark^\mp$ are forbidden.

The anomaly allows one to probe some details of the UV theory while only measuring a few IR processes.
The SU($N_f$) generators $T^a$ and SU(2)$_L$ generators $J^i$ in \cref{eq:Lanomaly} depend on $N_f$, which causes the strength of the anomaly interaction also to have dependence on $N_f$ that is captured by the coefficient $c$ in \cref{eq:c_value}.
In an idealized scenario, one could precisely measure the rate for some process involving self-interactions of the pNGBs to extract $f_\pi$, as well as the rates for the anomaly-induced processes to extract $c\,N_c/f_\pi$.
Then, one may be able to reconstruct the UV parameters $N_c$ and $N_f$.
(Since $N_c$ and $N_f$ are discrete, the product $c\,N_c$ is usually not degenerate between different combinations of $N_c$ and $N_f$.) 
This is analogous to using SM pion decay rates to extract the number of colors in QCD.

\subsection{Summary of Meson Properties}\label{sec:summary}

We have now explored each pNGB of chiral symmetry breaking and their decays. 
A major goal of this discussion was to elucidate the unique features of each meson, laying the groundwork for exploring their rich collider signals.
This section contains tables and figures that summarize the conclusions of the previous sections and serve as convenient references. 

In particular, the previous sections discussed the mass spectrum, decay channels, and collider production modes of the various mesons. Concerning the spectrum, we highlighted the model features that lift the degeneracy between different meson masses (see \cref{sec:mass} for details):
\begin{itemize}
    \item Electroweak loops induce quadratically divergent mass corrections $\sim J(J+1) \, f_\pi$ for mesons in SU(2)$_L$ representation $J$, lifting degeneracies and allowing higher multiplets to decay into lower ones with same G-parity. 

    \item Within an SU(2)$_L$ multiplet, electroweak loops induce mass splittings $\sim \alpha_W \, Q^2 \, m_W$, making lower-charge mesons lighter.

    \item The axial anomaly generates a mass contribution $\sim 4\pi \, f_\pi/N_c$ for the flavor-singlet $\etaDark^\prime$, which becomes negligible at large $N_c$. To avoid additional spectral complications, we focus on small $N_c$, where $\etaDark^\prime$ is heavy and decoupled, and neglect it in our collider analysis.
\end{itemize}

\begin{figure}[t]
\centering

\begin{minipage}[t]{0.23\textwidth}
  \vspace{0em}
  \includegraphics[width=\linewidth]{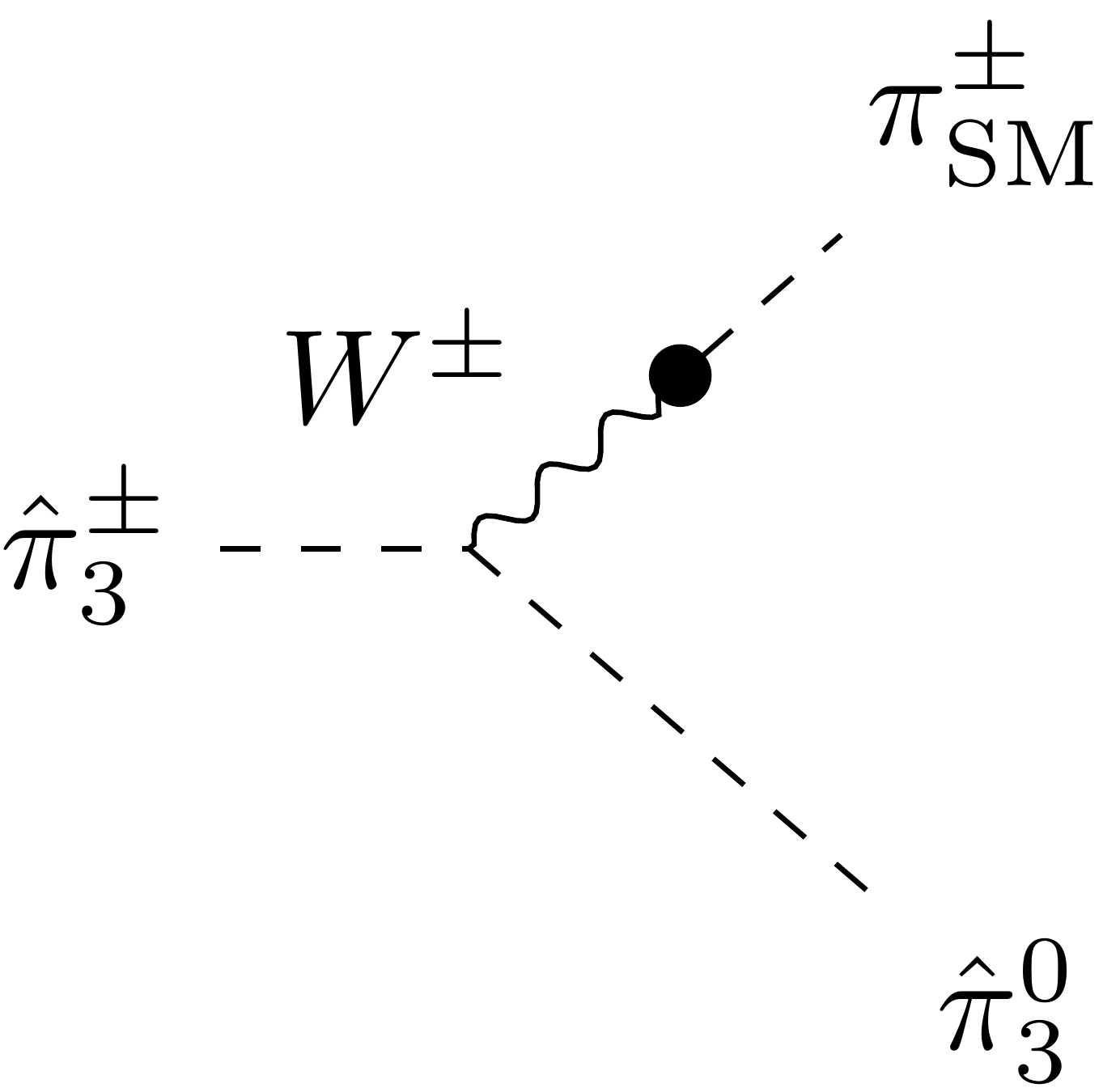}
\end{minipage}
\hspace{0.09\textwidth} 
\begin{minipage}[t]{0.18\textwidth}
  \vspace{1.8em}
  \includegraphics[width=\linewidth]{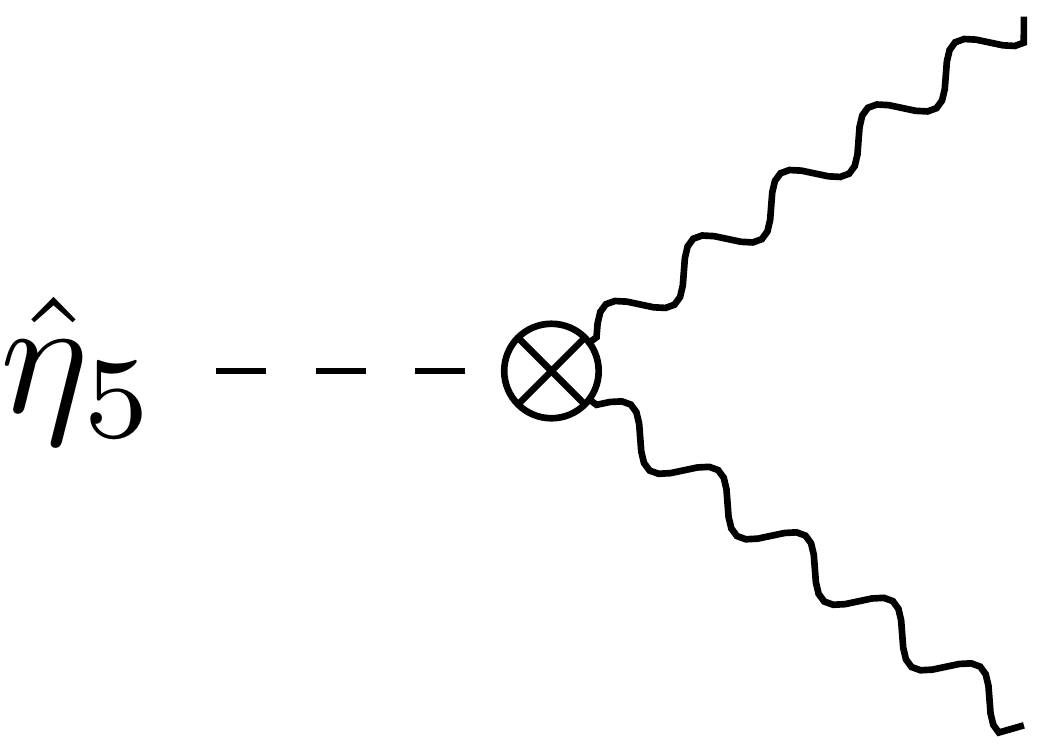}
\end{minipage}
\hspace{0.05\textwidth} 
\begin{minipage}[t]{0.3\textwidth}
  \vspace{-0.26em}
  \includegraphics[width=\linewidth]{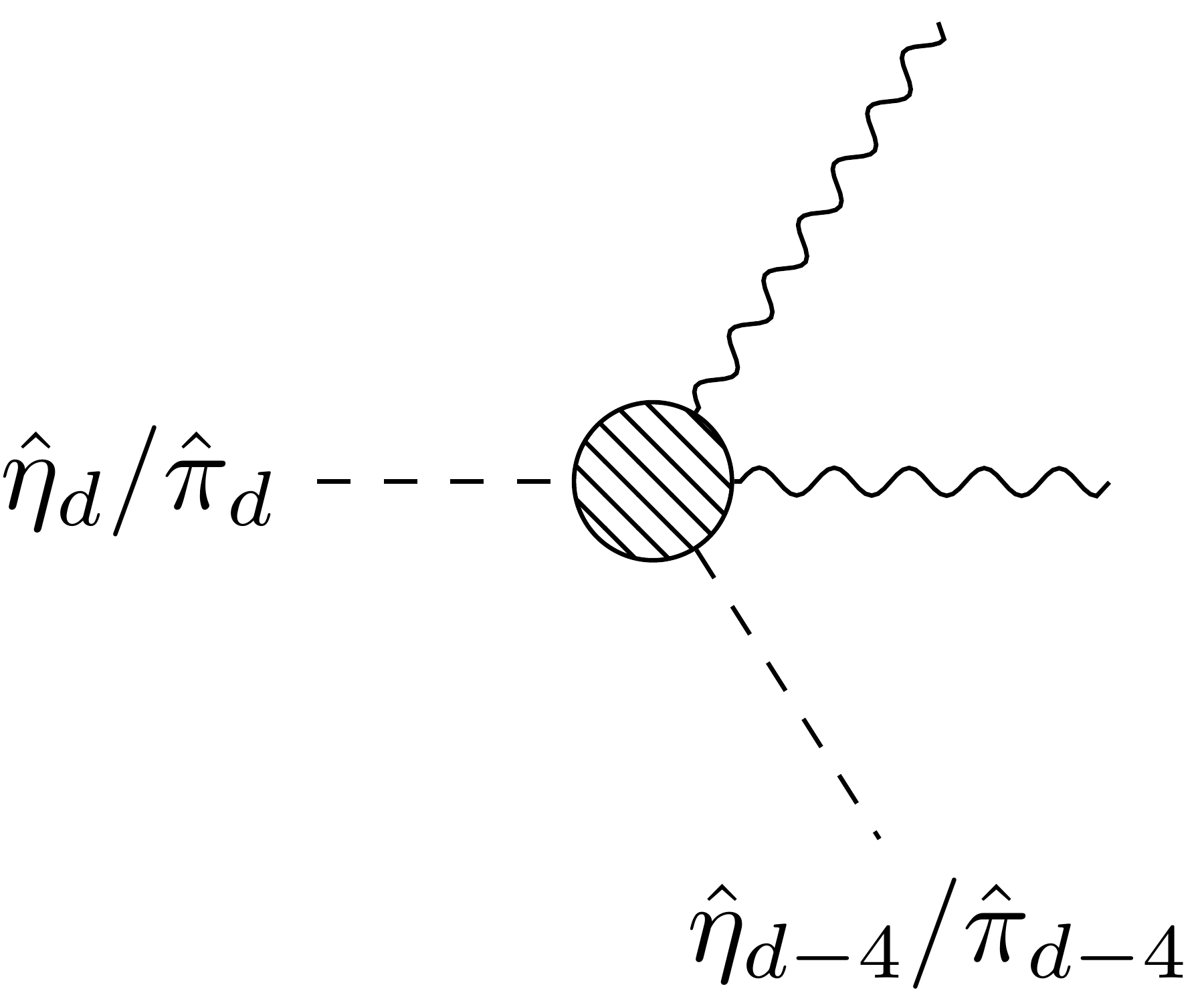}
\end{minipage}
\caption{\label{fig:decays} Diagrams depicting decay of the $\piDark_3^\pm$ via cascade (left), $\etaDark_5$ via the anomaly (middle), and higher multiplets via hopping (right). For the $\piDark_3^\pm$, the radiated off-shell $W$ boson effectively mixes with the SM pion. For the $\etaDark_5$, the decay products are pairs of electroweak bosons, and the $\otimes$ denotes insertion of the anomaly operator discussed in \cref{sec:anomaly}. For the higher multiplets, the meson hops down to a lower multiplet of the same $G$-parity by emitting two electroweak bosons, and the blob denotes insertion of the loop-induced vertex shown in \cref{sec:mass}.}
\end{figure}

These details of the mass spectrum directly feed into the study of different decay channels for various mesons. 
The relevant decay mechanisms are summarized in \cref{fig:decays} and in \cref{tab:decays}.
The decay of the lightest neutral meson $\piDark_3^0$ requires a breaking of the $G$-parity at scale $\Lambda_\slashed{G}$, see Eq.~\eqref{eq:pi30_lifetime}. 
For the $\piDark_3^\pm$ and $\etaDark_5$ mesons, the main decay channels are, respectively, the cascade decays (see Eq.~\eqref{eq:piRate} and the left diagram in \cref{fig:decays}) and the anomaly channel (see \cref{sec:anomaly} and the middle diagram in \cref{fig:decays}). 
Finally, $\piDark$ and $\etaDark$ mesons in higher SU(2)$_L$ multiplets decay to smaller representations via hopping (see Eq.~\eqref{eq:hop} and the right diagram in \cref{fig:decays}). 
The model parameters governing these channels are highlighted in \cref{tab:decays}. 


\begin{table}[t]
\centering
\begin{tabular}{c|c|c}
\hline
Species & Dominant Decay Mechanism & Relevant Parameters \\ \hline
$\hat{\pi}_3^0$ & $G$-parity violation & $\Lambda_{\slashed{G}}$ \\ \hline
$\hat{\pi}_3^\pm$ & cascade & SM pion mass and decay constant \\ \hline
$\hat{\eta}_5$ & anomaly & $N_c$, $N_f$, $f_\pi$ \\ \hline
$\hat{\pi}_7$, $\hat{\eta}_9$, $\hat{\pi}_{11}$, \ldots & hopping & $J$, $f_\pi$
\\
\hline
\end{tabular}
\caption{\label{tab:decays} Dominant decay mechanisms of each pNGB of dark chiral symmetry breaking and the theory parameters (besides the masses of the mesons themselves) that determine the decay rates. The cascade, anomaly, and hopping mechanisms are pictured in \cref{fig:decays}, and the dark $G$-parity violation would stem from some UV completion (with $\Lambda_{\slashed{G}}$ the associated UV scale) that is unspecified in this work.}
\end{table}

Finally, in \cref{tab:summary} we gather the relevant properties of $\piDark_3$ and $\etaDark_5$ mesons, including their main production channel at LHC. For the remainder of this work, we focus on these particular mesons and their collider signatures. 


\begin{table}[t]
\centering
\begin{tabular}{c|c|c|c}
\hline
Species & Production & Lifetime & Signal \\
\hline
$\piDark_3^\pm$ & pairs via & $\mathcal{O}(0.1)\,\text{ns}$  & \multirow{2}{*}{disappearing tracks} \\
\cline{1-1}
\cline{3-3}
$\piDark_3^0$ & DY and VBF & collider-stable & \\
\hline
\hline
\multirow{2}{*}{$\etaDark_5^{0}$, $\etaDark_5^{\pm}$, $\etaDark_5^{\pm\pm}$} & resonances via & \multirow{2}{*}{prompt} & \multirow{2}{*}{diboson decays} \\
 & the anomaly & & \\
\hline
\end{tabular}
\caption{\label{tab:summary} Summary of the dark mesons we consider in our collider study in \cref{sec:signals}.
}
\end{table}

\section{Collider Signals and Constraints}
\label{sec:signals}

The results of \cref{sec:model} imply striking collider signals of the model's dark mesons. 
Building on that, in this section we study the signatures of the two lightest meson representations $\piDark_3$ and $\etaDark_5$. 
The $\piDark_3$ and $\etaDark_5$ multiplets have entirely distinct and complementary signatures, each of which is the subject of several LHC searches. We use these searches to place constraints on the meson masses and the ratio $f_\pi/N_c$.  For event generation, we use \textsc{MadGraph5\_aMC@NLO} \cite{Alwall:2014hca,Frederix:2018nkq} with a model we implement in \textsc{FeynRules} \cite{Cacciari:2005hq,Cacciari:2011ma}. For the LLP study, we use \textsc{Pythia} 8 \cite{Bierlich:2022pfr} for showering and \textsc{Delphes} 3 \cite{deFavereau:2013fsa} with the default ATLAS card for detector simulation, which includes jet clustering using \textsc{FastJet} \cite{Cacciari:2005hq,Cacciari:2011ma} with the anti-$k_T$ clustering algorithm \cite{Cacciari:2008gp}. We generate events with zero, one, or two jets in the hard process, which are combined by \textsc{MadGraph5} with matrix-level matching.

\subsection{Disappearing Tracks: A 3-plet Signature}
\label{sec:tracks}

The $\piDark_3^\pm$ decays with a lifetime of $\mathcal{O}(0.1)\,\text{ns}$ to a neutral collider-stable $\piDark_3^0$ and a soft SM pion, which leads to the same kind of signal as the classic chargino decay to a neutralino in the Minimal Supersymmetric Standard Model when the wino is much lighter than all of the other superpartners. The charged particle leaves a few hits in the inner tracker before it decays, the soft SM pion is not reconstructed, and the result is a track that seems to stop inside the tracker.\footnote{There has been progress in reconstructing the soft SM pions that have $p_T\gtrsim 500\,\text{MeV}$, which helps resolve the decay vertex and reject fake tracks that come from random hits \cite{ATLAS:2019pjd}.} These disappearing tracks signals are the subjects of recent searches by the ATLAS and CMS collaborations \cite{ATLAS:2022rme,CMS:2023mny}, which place strong lower bounds on chargino masses. 

\begin{figure}[t]
\centering
\begin{minipage}[t]{0.21\textwidth}
  \vspace{0em}
  \includegraphics[width=\linewidth]{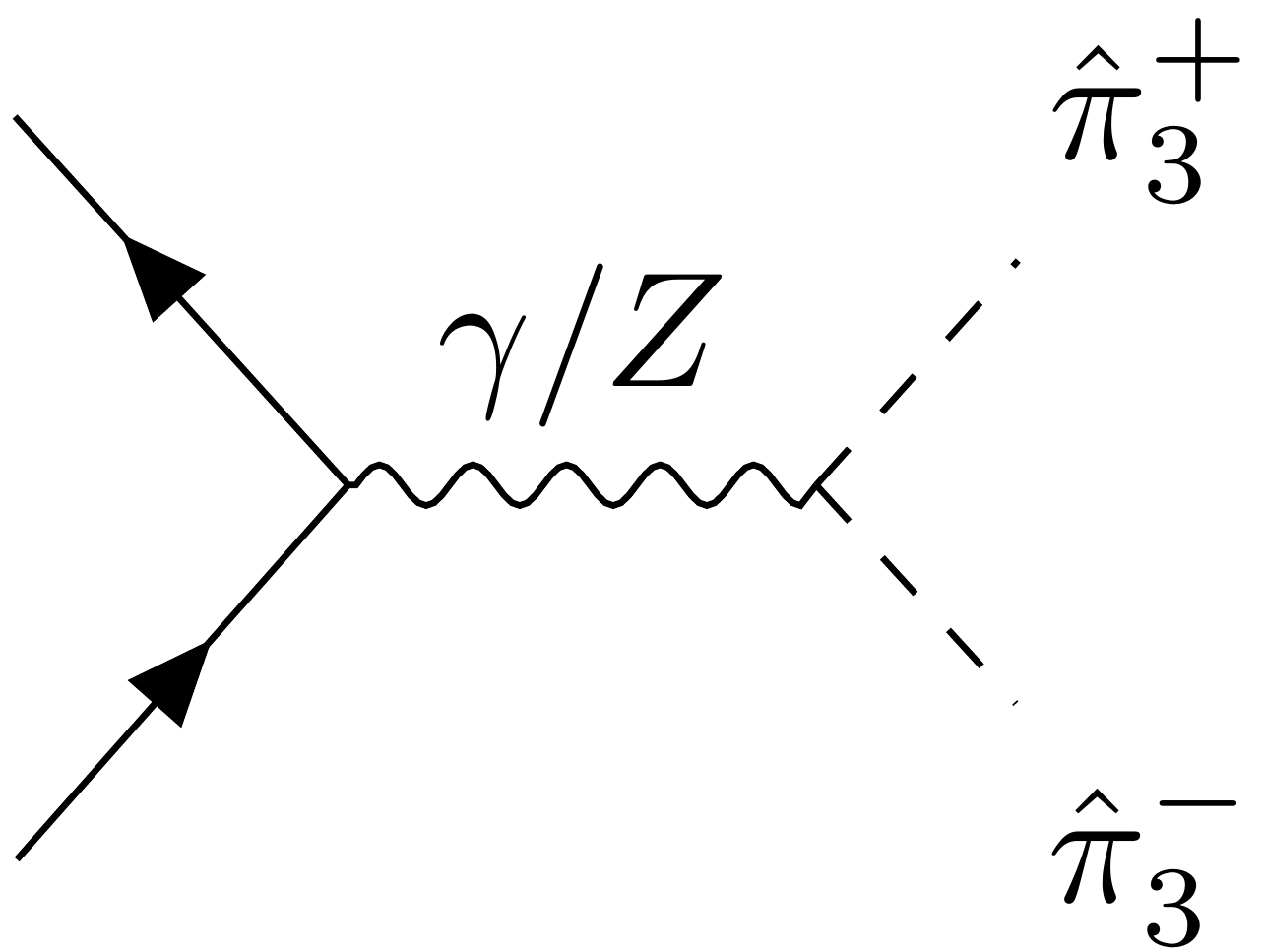}
\end{minipage}
\hspace{0.05\textwidth} 
\begin{minipage}[t]{0.21\textwidth}
  \vspace{0em}
  \includegraphics[width=\linewidth]{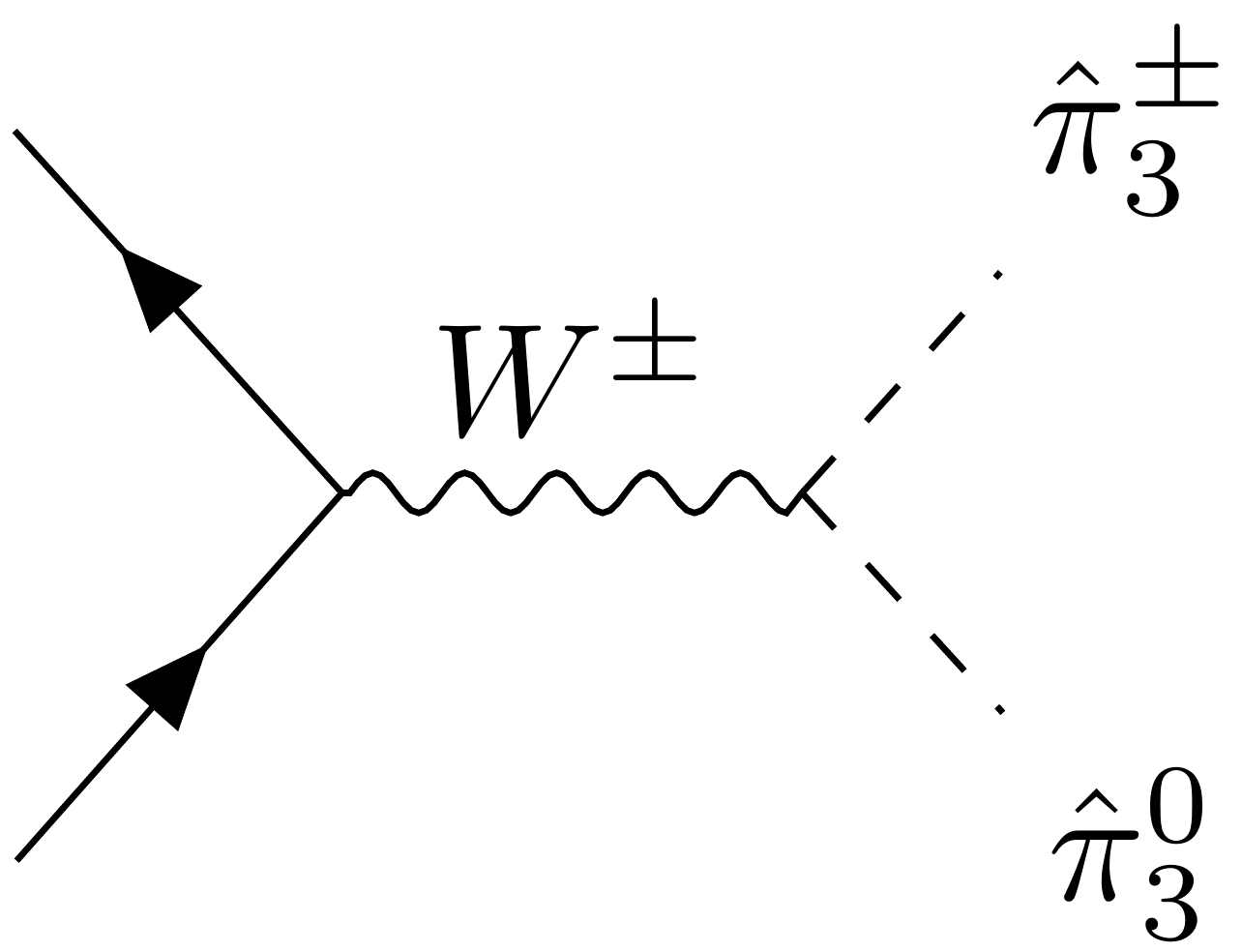}
\end{minipage}

\begin{minipage}[t]{0.25\textwidth}
  \vspace{0em}
  \includegraphics[width=\linewidth]{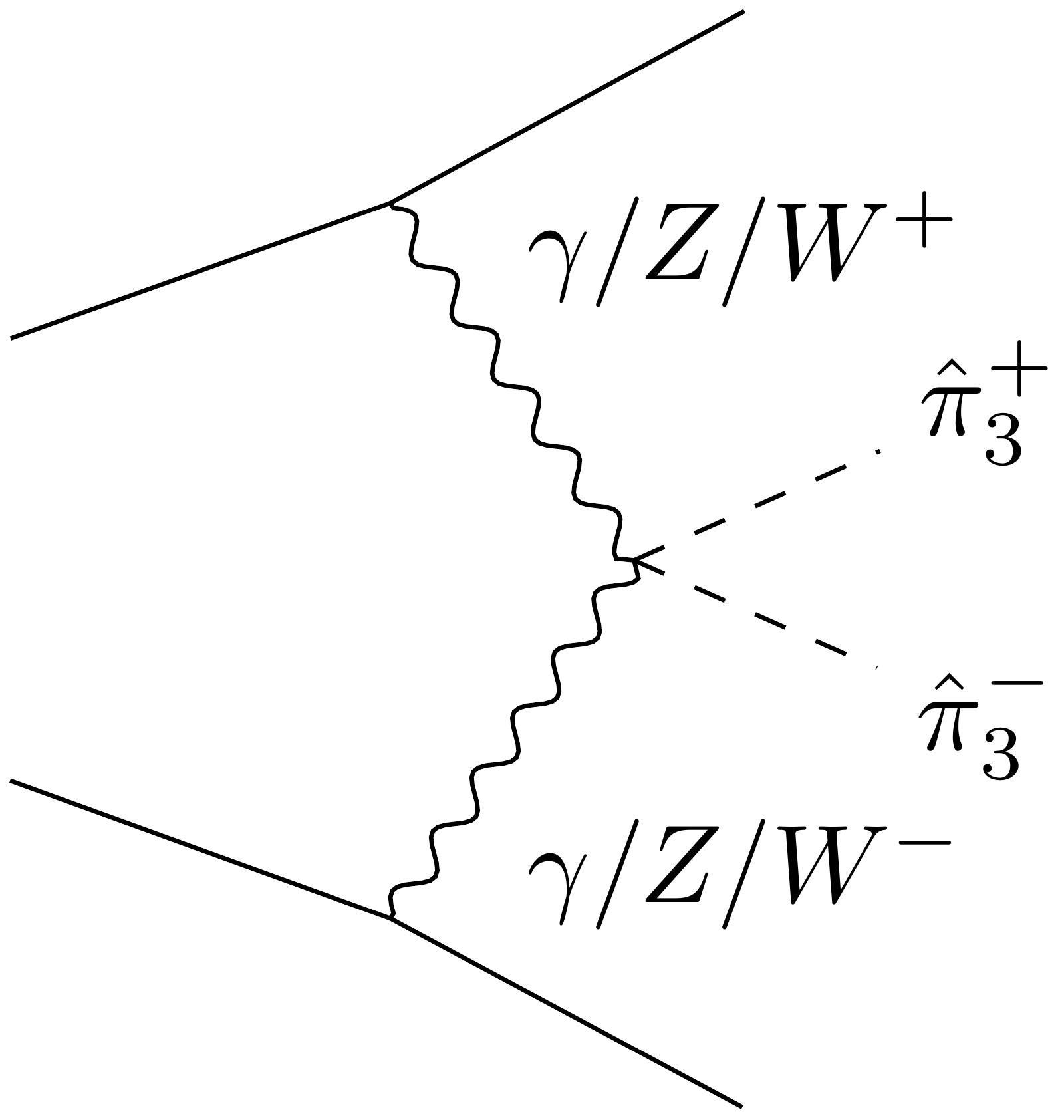}
\end{minipage}
\hspace{0.01\textwidth} 
\begin{minipage}[t]{0.25\textwidth}
  \vspace{0em}
  \includegraphics[width=\linewidth]{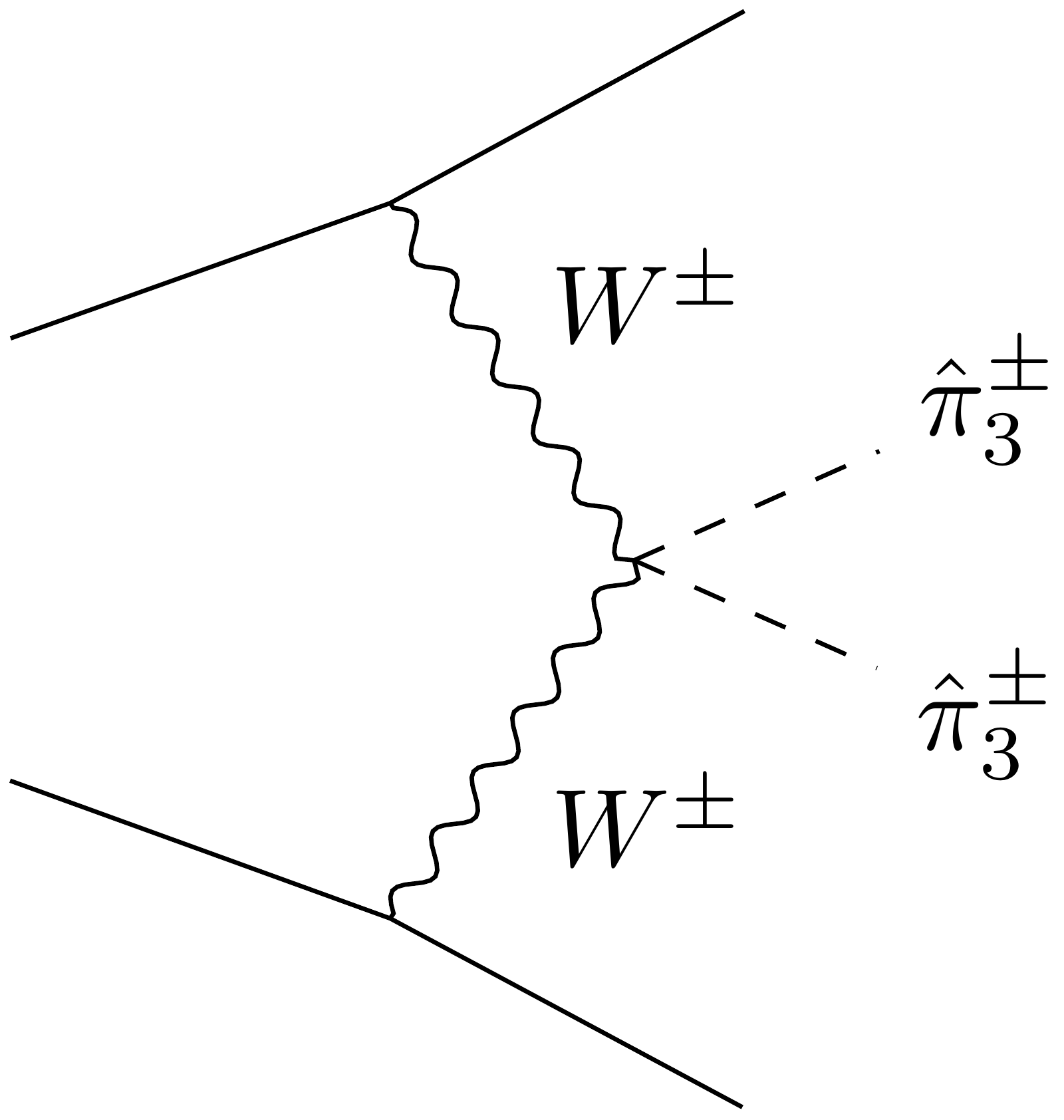}
\end{minipage}
\hspace{0.01\textwidth} 
\begin{minipage}[t]{0.25\textwidth}
  \vspace{0em}
  \includegraphics[width=\linewidth]{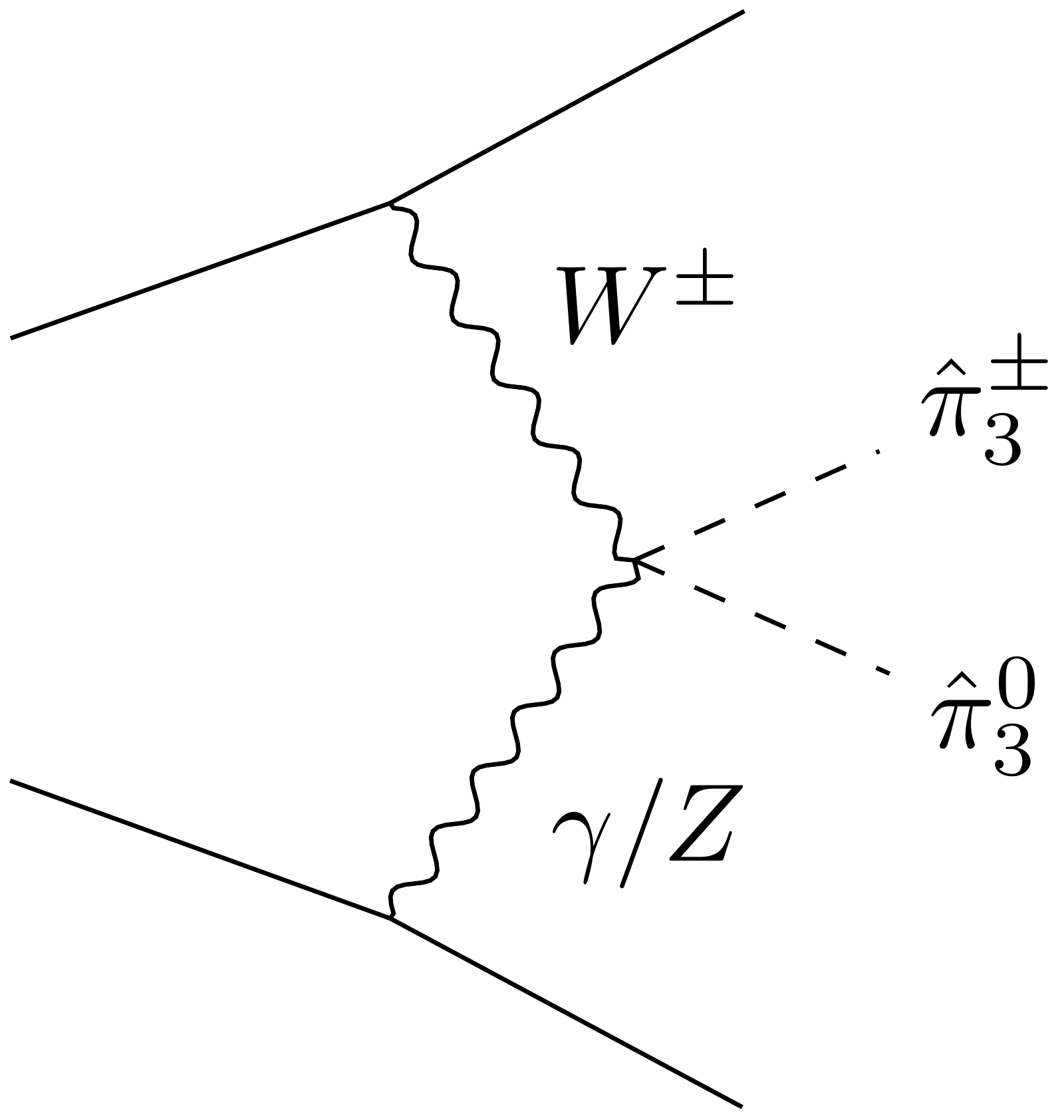}
\end{minipage}
\caption{\label{fig:pi3production} Diagrams representing pair production of the $\piDark_3$ mesons via Drell-Yan (top) and vector boson fusion (bottom). The VBF processes also get contributions from diagrams with three-point vertices coupling the mesons to vector bosons. Diagrams with  production of two neutral mesons via VBF are not pictured, as we do not consider such processes.}
\end{figure}

\begin{figure}[t]
  \centering
  \includegraphics[width=0.81\textwidth]{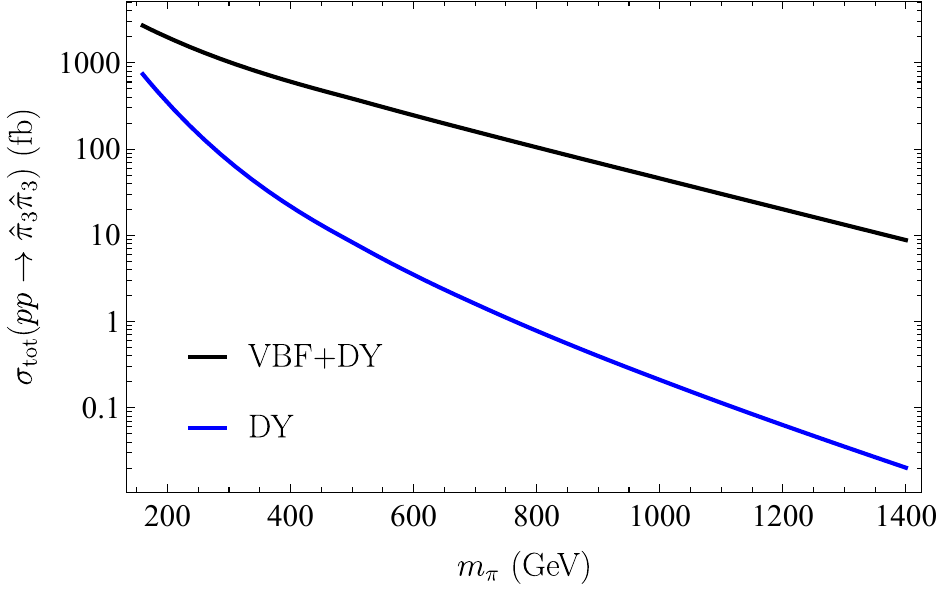} 
  \caption{\label{fig:VBFDY} Total LHC production cross sections of the dark meson triplet $\piDark_3$ pairs we consider. The production rate for exclusively Drell-Yan processes falls much faster as a function of the $\piDark_3$ mass $m_\pi$ than the rate of processes that include vector boson fusion.}
\end{figure}

The $\piDark_3$ is produced via DY and VBF in opposite charge, same charge, or charged-plus-neutral pairs, as pictured in \cref{fig:pi3production}. As has been pointed out in the contexts of sleptons and charginos \cite{Datta:2001hv,Choudhury:2003hq}, the VBF cross section falls slower than that of DY as a function of mass, thus it can become the dominant process for masses above an $\mathcal{O}(100)\,\text{GeV}$ threshold. This effect is partially due to DY suffering from an $s$-channel propagator, while the vector bosons facilitating VBF can be light and close to on-shell. Another contribution to the enhancement of VBF over DY is that DY requires a sea anti-quark in the initial state, which has a suppressed parton distribution function compared to the quark initial states allowed in VBF. As shown in \cref{fig:VBFDY}, we find that VBF dominates DY by a factor of $\mathcal{O}(100)$ for $m_\pi\gtrsim700\,\text{GeV}$ in our case. This is a stronger enhancement than, for example, that of a selectron, in part due to the combinatorics involved in producing the various states within the $\piDark_3$ multiplet. The large production cross section allows us to place even stronger constraints on the $\piDark_3$ mass than existing constraints on chargino masses.

If $N_f\geq4$, then there is a $\piDark_7$ in the meson spectrum in addition to the $\piDark_3$. The decays of the $\piDark_7$ to the $\piDark_3^\pm$ may also contribute to the disappearing tracks signal. However, these hopping decays (hopping from heavier to light pions with the same $G$-parity) are associated with additional SM states due to the emitted EW bosons, which complicates the disappearing tracks analysis. Therefore, we do not include the $\piDark_7$ (or larger multiplets present at higher $N_f$) in our analysis.

\begin{figure}[t]
  \centering
  \includegraphics[width=0.77\textwidth]{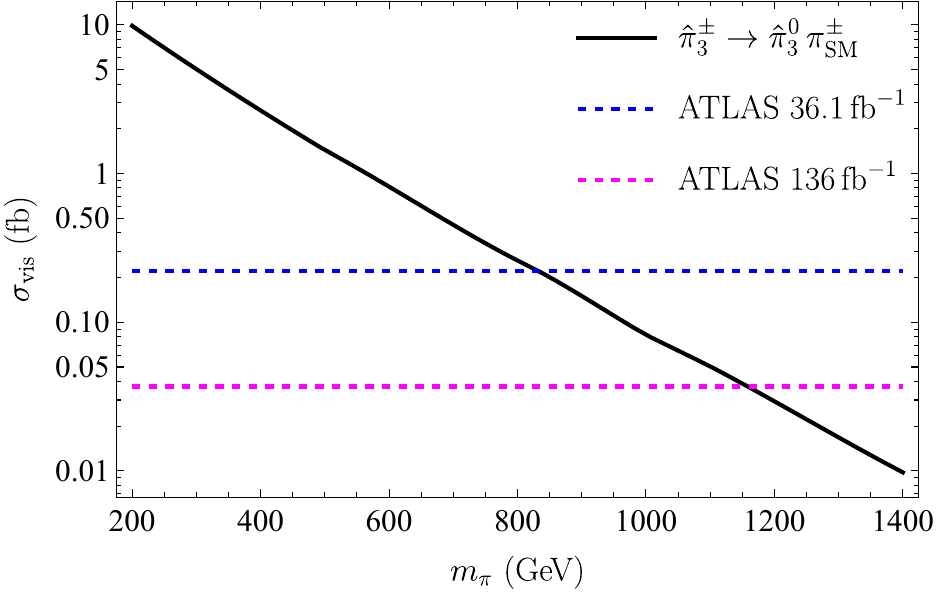} 
  \caption{\label{fig:trackConstraint} The visible cross section for production of disappearing tracks due to decays of the long-lived $\piDark_3^\pm$ meson. The dashed lines indicate model-independent upper bounds on this cross section based on the ATLAS searches in Refs.~\cite{ATLAS:2017oal,ATLAS:2022rme} and using the re-interpretation methods provided by Ref.~\cite{Belyaev:2020wok}. This shows a strong lower bound on the $\piDark_3$ meson mass $m_\pi$ of $\sim\!1.2\,\text{TeV}$.}
\end{figure}

We re-use the methods from Ref.~\cite{Belyaev:2020wok} available in the LLP Recasting Repository \cite{Cottin:2019llp} to re-interpret two Run 2 ATLAS disappearing tracks searches.\footnote{The most recent ATLAS and CMS searches in Refs.~\cite{ATLAS:2022rme,CMS:2023mny} derived very similar model-dependent constraints on masses of supersymmetric particles, so we expect the CMS constraints on our model to be closely comparable.} The reinterpretation package enables us to compute the visible cross section, \emph{i.e.}~the product of the total production cross section and the efficiency of identifying a disappearing track. The first search, which the re-interpretation code was built for, used a dataset with an integrated luminosity of $36.1\,\text{fb}^{-1}$ and placed a model-independent constraint on the visible cross section of disappearing tracks of $0.22\,\text{fb}$ at the 95\% confidence level \cite{ATLAS:2017oal}.  The second search used an integrated luminosity of $136\,\text{fb}^{-1}$ and found a constraint of $0.037\,\text{fb}$ \cite{ATLAS:2022rme}. Our use of the more recent search limit on the visible cross section while taking advantage of the previous search reinterpretation methods provides a rough estimate that assumes the two searches have similar signal efficiencies. In \cref{fig:trackConstraint} we show the visible cross section computed in our model as a function of the $\piDark_3$ mass. The efficiency is $\mathcal{O}(0.1)\%$ across the mass range in the figure. We find that the more recent search constrains the $\piDark_3$ mass to be $m_\pi \gtrsim 1.2\,\text{TeV}$. 

The $\piDark_3$ production rate is independent of $f_\pi$, $N_c$, and $N_f$ (unlike the resonance signature discussed in the upcoming section). 
This result therefore imposes a remarkably strong lower bound on the masses of all mesons in the spectrum, particularly since the $\piDark_3$ is the lightest state. Recall from \cref{sec:mass} that the relationship between the $\piDark_3$ mass and the $\etaDark_5$ mass $m_\eta$ is somewhat uncertain due to its dependence on the UV cutoff of the meson EFT. Therefore, we cannot directly translate the bound on $m_\pi$ into a bound on $m_\eta$, but we can estimate based on the scaling discussed in \cref{sec:mass} that $m_\eta-m_\pi$ is roughly within an $\mathcal{O}(1)$ factor of $f_\pi$. This places an even larger lower bound on $m_\eta$ that complements the constraints we show in \cref{sec:resonances}.

\subsection{Anomaly-induced Resonances: A 5-plet Signature}
\label{sec:resonances}

The $\etaDark_5$ mesons can be singly produced through VBF and dominantly decay to pairs of electroweak bosons via the anomaly explained in \cref{sec:anomaly} and shown in \cref{fig:anomaly}. 
Recall that while the $\piDark_3$ exists in the spectrum for all $N_f\geq2$, the $\etaDark_5$ is only present for $N_f\geq3$. As a concrete benchmark, we set $N_f=4$ in this section, but changing $N_f$ simply modifies cross sections by $\mathcal{O}(1)$ factors explained in \cref{eq:5pletAnomaly,eq:c_value}. Setting $N_f=4$ is a motivated choice, since it results in larger production rates than $N_f=3$, and having $N_f\geq5$ often leads to nearby Landau poles in the SU(2)$_L$ running coupling \cite{Asadi:2024tpu}.

\begin{figure}[t]
  \centering
  \includegraphics[width=0.73\textwidth]{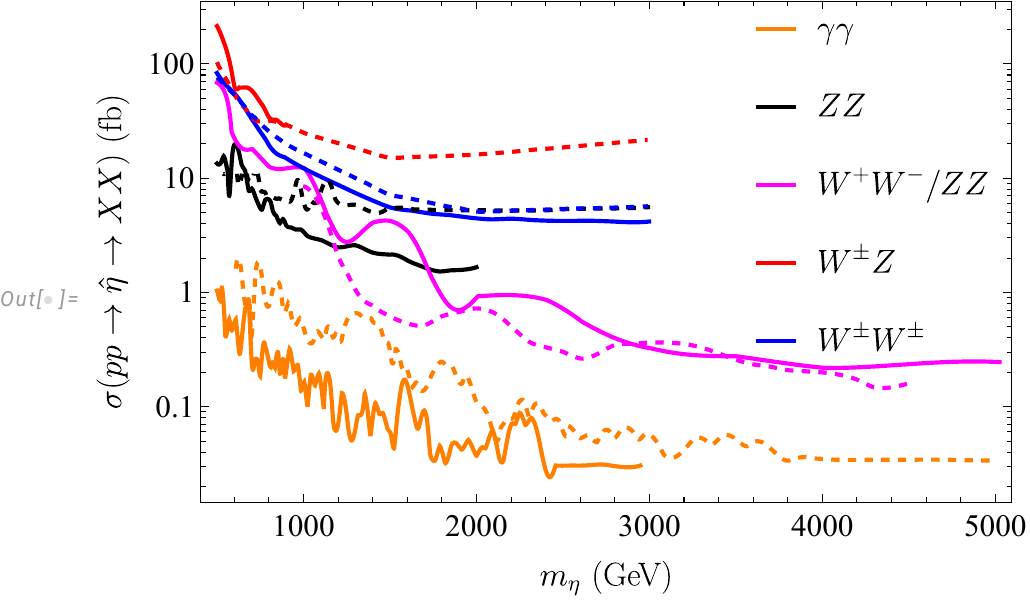} 
  \caption{\label{fig:limits} Upper bounds on the cross section to produce various final states from the searches in Refs.~\cite{ATLAS:2018iui,ATLAS:2020tlo,ATLAS:2020fry,CMS:2021wlt,CMS:2021klu,ATLAS:2021uiz,ATLAS:2023sua,CMS:2024nht,CMS:2024vps} through resonant production and decay of the scalar $\etaDark$. Solid (dashed) lines correspond to ATLAS (CMS) searches. The $ZZ$ searches used the fully leptonic channel, and the $W^+W^-/ZZ$ searches used the semileptonic channel. In the context of our model, the $\etaDark$ is one of the pNGBs of dark chiral symmetry breaking in the SU(2)$_L$ 5-plet. Besides the diphoton searches, each of these is specialized to production via VBF. We interpret these upper bounds as lower bounds on $4\pi f_\pi/N_c$ in \cref{fig:fPiLimits}.}
\end{figure}

There are many LHC searches for resonances decaying to two electroweak bosons. Various final states have been exploited, including diphoton \cite{ATLAS:2021uiz,CMS:2024nht}, $ZZ$ in the fully leptonic channel \cite{ATLAS:2020tlo,CMS:2024vps}, $W^+W^-$ and/or $ZZ$ in the semileptonic channel \cite{ATLAS:2020fry,CMS:2021klu}, $W^\pm Z$ \cite{ATLAS:2018iui,CMS:2021wlt}, and same sign $W$ pairs \cite{CMS:2021wlt,ATLAS:2023sua}. 
Each of these searches quotes limits on the total production cross sections of a scalar resonantly produced via VBF and decaying to each pair of bosons (except the diphoton searches, which do not specify the production channel).\footnote{The diphoton search in Ref.~\cite{ATLAS:2021uiz} quotes a limit on the fiducial cross section, which is defined in terms of generator-level cuts. These cuts modify our cross sections only by a small $\mathcal{O}(1)$ factor.}

We show the upper limits on the cross sections from each search in \cref{fig:limits}.\footnote{The searches in Refs.~\cite{ATLAS:2016mti,CMS:2016ssv} target the $\gamma Z$ final state, which is a principal decay mode of the $\etaDark_5^0$, but these specialize to the gluon-gluon fusion production channel, so we do not include them.} 
While the diphoton searches probe the lowest cross sections, the branching ratio of $\etaDark_5^0\to\gamma\gamma$ is about $4\%$, and the branching ratios to $W^+W^-$ and $ZZ$ are each roughly $1/3$. Likewise, the same sign $W$ searches tend to probe higher cross sections than other final states, but this is the principal decay mode of the $\etaDark_5^{\pm\pm}$. Therefore, it is not obvious \textit{a priori} which search imposes the strongest constraint on the model, and future searches for final states that currently impose weaker bounds may become competitive with searches for other final states if they attain greater signal efficiency. 

To get a sense of the $\etaDark_5$ decay widths, we can use the partial width

\begin{equation}
\begin{aligned}
    \Gamma(\etaDark_5^0\to \gamma \gamma) &= \frac{2}{\pi}\alpha^2\, m_\eta^3\left( \frac{N_c}{4\pi f_\pi} \right)^2 \\
    &\simeq 0.039\,\text{GeV}\left( \frac{m_\eta}{1\,\text{TeV}} \right)^3 \left( \frac{1\,\text{TeV}}{4\pi f_\pi/N_c} \right)^2,
\end{aligned}
\end{equation}

\noindent where $\alpha\simeq 1/128$ is the electromagnetic coupling, and $N_f=4$. Since the branching ratio to diphoton is $\sim\!4\%$, the total widths $\Gamma_\eta$ are $\mathcal{O}(1)\,\text{GeV}$ when the other relevant mass scales are $\mathcal{O}(1)\,\text{TeV}$. Likewise, the ratio $\Gamma_\eta/m_\eta$ is $\mathcal{O}(10^{-3})$, so the narrow width approximation is valid in this regime.

\begin{figure}[t]
  \centering
  \includegraphics[width=0.98\textwidth]{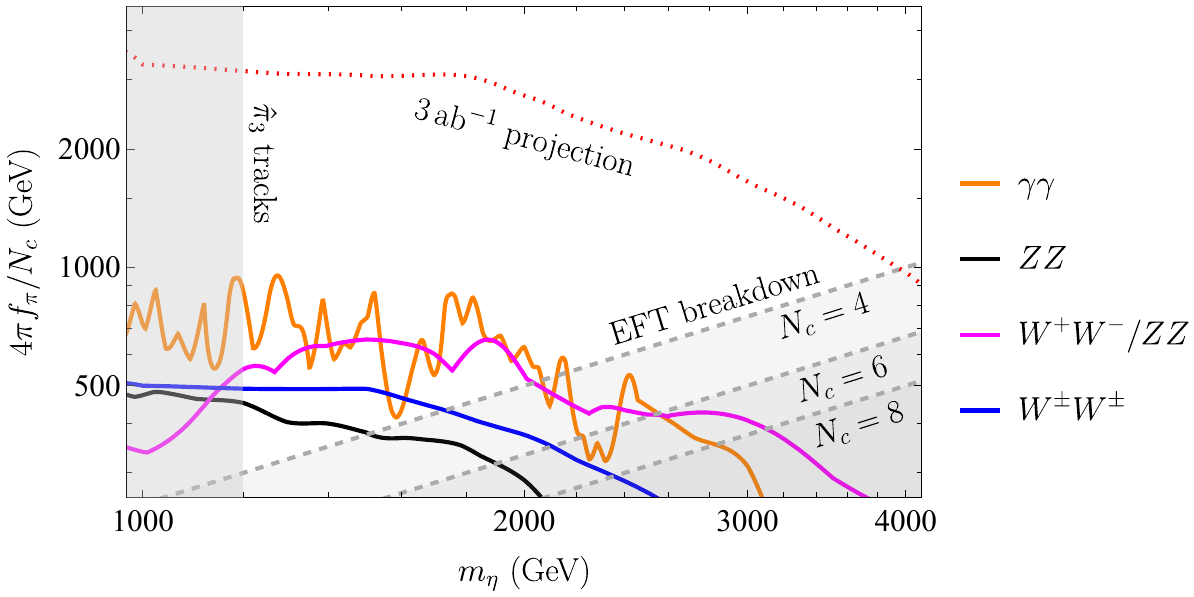}
  \caption{\label{fig:fPiLimits} Lower bounds on the ratio $4\pi f_\pi/N_c$ as a function of the $\etaDark_5$ meson mass $m_\eta$ using the strongest constraints in \cref{fig:limits} for each diboson final state. The gray contours indicate lower bounds on $4\pi f_\pi/N_c$ for a few values of $N_c$, below which $m_\eta>4\pi f_\pi$ and the EFT of mesons breaks down. The region with $m_\eta<1.2\,\text{TeV}$ is excluded by the disappearing tracks analysis in \cref{sec:tracks}.  The dotted red contour indicates a projection from a na\"{i}ve luminosity rescaling of the expected limits of the searches with the strongest constraints on our model for a $3\,\text{ab}^{-1}$ dataset.}
\end{figure}

\begin{figure}[t]
  \centering
  \includegraphics[width=0.77\textwidth]{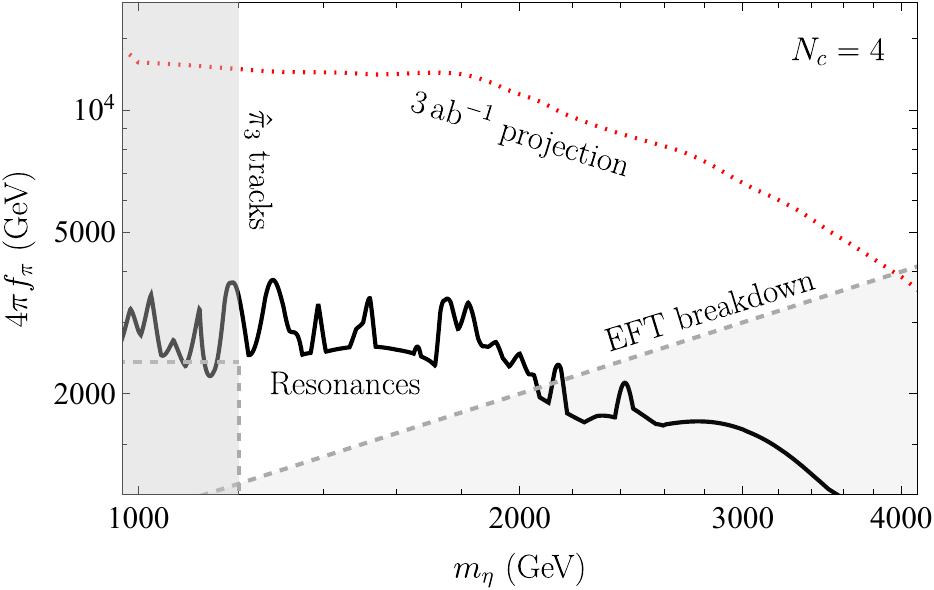} 
  \caption{\label{fig:fPiNc4}Lower bounds on $4\pi f_\pi$ as a function of the $\etaDark_5$ meson mass $m_\eta$ in the case of $N_c=4$. Existing and projected resonance constraints due to the 5-plet anomaly, constraints from disappearing tracks due to the dark meson 3-plet $\piDark_3$, and the meson EFT breakdown where $m_\eta>4\pi f_\pi$ are shown as in \cref{fig:fPiLimits}. The disappearing tracks constraint below $4\pi f_\pi=2.4\,\text{TeV}$ is indicated by a dashed line, where pair production of $\piDark_3$ mesons with a mass of $1.2\,\text{TeV}$ is outside the EFT regime of validity.}
\end{figure}

In \cref{fig:fPiLimits}, we use the fact that the cross section scales as $(N_c/f_\pi)^2$ to recast the upper bounds on the production cross sections in \cref{fig:limits} as lower bounds on the ratio $4\pi f_\pi/N_c$ as a function of $m_\eta$. The searches for diphoton and $W^+W^-/ZZ$ in the semileptonic channel impose very similar constraints, and the same sign $W$ and fully leptonic $ZZ$ constraints are only an $\mathcal{O}(1)$ factor below. Constraints due to $W^\pm Z$ searches are not shown, as they are too weak to exclude any valid parameter space. 

While we cannot precisely compute $m_\eta-m_\pi$, we know $m_\eta>m_\pi$, so the disappearing tracks analysis in \cref{sec:tracks} excludes the region where $m_\eta\lesssim1.2\,\text{TeV}$. The figure also shows contours where $m_\eta = 4\pi f_\pi$ for a few values of $N_c$. For values of $f_\pi/N_c$ below these contours, one is probing the theory at a scale outside of the EFT's regime of validity, so that range of parameters is not physical. Our comparison of the $\piDark_3$ constraint with the $\etaDark_5$ constraints presents another possible issue of EFT validity. The $\piDark_3$ pair production mechanisms we consider occur via marginal operators, so we can always implicitly enforce in \cref{fig:trackConstraint} that the EFT scale $4\pi f_\pi$ is greater than the invariant mass of the $\piDark_3$ pairs. However, for smaller values of $4\pi f_\pi$ in \cref{fig:fPiLimits}, \textit{pairs} of $\piDark_3$ mesons with masses of $1.2\,\text{TeV}$ may have a greater invariant mass than the EFT scale, in which case the disappearing tracks constraint does not have a sensible interpretation within the EFT. This issue is most pronounced for smaller values of $N_c$.

In \cref{fig:fPiNc4}, we show the anomaly-induced resonance constraints for the choice of $N_c=4$ and find that the bound on $4\pi f_\pi$ is greater than 2\,TeV for the full $\etaDark_5$ mass range within the EFT's regime of validity. 
Recall from \cref{sec:mass} that heavier meson resonances may become visible when $4\pi f_\pi\lesssim2\,\text{TeV}$. 
The figure shows that the anomaly-induced resonance signatures are at least competitive with those of the vector mesons, and they may be dominant. However, this 2\,TeV estimate is predicated on a loose analogy between our model and the dark sector studied in Refs.~\cite{Kribs:2018oad} and \cite{Kribs:2018ilo}. The actual mass threshold where vector resonances gain importance depends strongly on details of the model such as $N_c$, the $\rhoDark/\piDark$ mass ratio, and the branching ratios of the $\rhoDark$ decays to other meson species. A thorough analysis of the vector meson phenomenology is beyond the scope of this work.

The constraints we derived in this section apply to meson mass scales up to multiple TeV and depend on $f_\pi$ and $N_c$, while the disappearing tracks constraints in \cref{sec:tracks} apply to mass scales of $\sim\!1\,\text{TeV}$ and do not depend (at least explicitly) on $f_\pi$ or $N_c$. These two strategies are therefore complementary and represent exciting opportunities for future searches. 

In \cref{fig:fPiLimits,fig:fPiNc4}, we also show projected bounds from a na\"{i}ve luminosity rescaling of the expected limits from the searches with the strongest constraints on $4\pi f_\pi/N_c$. Specifically, we take the expected constraint on the cross section from each search, compute the corresponding constraints on $4\pi f_\pi/N_c$, and re-scale these constraints by $\sqrt{3\,\text{ab}^{-1}/L}$ with $L$ each search's luminosity. We then take the maximum of these re-scaled constraints for each value of $m_\eta$ as the projected constraint. This projection demonstrates that if future searches attain similar signal efficiencies to existing searches, and if the High-Luminosity LHC acquires a dataset of $3\,\text{ab}^{-1}$, then such searches could reach a significantly larger region of parameter space.

\section{Conclusions}
\label{sec:conclusion}

This paper studied dark mesons arising in the confined phase of a strongly-coupled sector with vector-like dark quarks.  
We chose the dark quarks to be in the fundamental representation of a new confining SU($N_c$) and the $N_f$-dimensional representation of SU(2)$_L$. 
Considering the limit where the dark quark mass is much less than the dark confinement scale, we discussed the spectrum of SU(2)$_L$ representations of these mesons, their masses, and their transformations under various $\mathbb{Z}_2$ symmetries including a dark $G$-parity. 
We then studied two exotic collider signatures of the dark mesons in detail:
\begin{itemize}
    \item The $G$-odd mesons are long-lived and therefore excellent targets for current and future LLP searches. In particular, the singly charged 3-plet $\piDark_3^\pm$ give rise to disappearing track signals.
    \item If $N_f\geq3$, there is a $G$-even 5-plet meson $\etaDark_5$ in the spectrum, which we showed is the unique pNGB of the dark sector chiral symmetry breaking that has an anomaly with SU(2)$_L$. This anomaly enables the $\etaDark_5$ to be resonantly produced through VBF and decay to pairs of electroweak bosons with a rate that depends on $N_c$ and the dark pion decay constant $f_\pi$.
\end{itemize}

We used the reinterpretation framework from Ref.~\cite{Belyaev:2020wok} and the disappearing tracks searches in Refs.~\cite{ATLAS:2017oal} and \cite{ATLAS:2022rme} to place a strikingly strong lower bound on the $\piDark_3$ mass $m_\pi$ of $\sim\!1.2\,\text{TeV}$. We also used the searches for scalar resonances decaying to electroweak boson pairs in Refs.~\cite{CMS:2021wlt,ATLAS:2020tlo,ATLAS:2020fry,CMS:2021klu,ATLAS:2021uiz,ATLAS:2023sua,CMS:2024nht,CMS:2024vps} to derive lower bounds on the ratio $4\pi f_\pi/N_c$ as a function of the $\etaDark_5$ mass $m_\eta$. 
Measuring the rates of the anomaly-induced processes could help reconstruct $N_c$ and $N_f$, thereby revealing a great deal about the UV theory while only measuring states in the IR. 
These anomaly-induced resonances therefore remain an exciting target for diboson searches, including those using the unusual same sign $W$ final state.

The model under study is an example of a confining dark sector with $\mathcal{H}$-parity \cite{Asadi:2024bbq}. The baryons in such models may be viable dark matter candidates that are characteristically challenging to probe via direct detection despite having constituents that are electrically charged. This is particularly true in the Noble Dark Matter case, where the dark matter candidate is either a total SM singlet or a singlet with a small mixing with other states \cite{Asadi:2024tpu}. As discussed in Refs.~\cite{Asadi:2024bbq,Asadi:2024tpu}, there are various opportunities for further analyzing the early universe dynamics, indirect detection, and astrophysical signatures of these dark matter scenarios. The results of this work show that a broad space of parameters with light 3-plet mesons or collider-scale 5-plet mesons is already ruled out, and future analyses can focus on other cases.

Several natural extensions of this work remain to be explored.
While $G$-parity may stabilize some of the meson species on a detector timescale, the possibility of $G$ violation by Planck-suppressed dimension-5 operators threatens the model's cosmological safety unless $G$ is broken at a sub-Planckian scale that shortens the lifetime to below the timescale of Big Bang Nucleosynthesis. The lifetime could still be well beyond the detector scale, which is the case we consider in this work, but interesting new signals may emerge if the neutral $\piDark$ mesons can decay within the detector. For example, the $\piDark_3^\pm$ could leave a track that disappears within the tracker but points to a displaced energy deposit within the calorimeter where the $\piDark_3^0$ decays later. Measuring this process would probe the scale of $G$ violation, and we leave the analysis of such signals for future work.
Likewise, decays of pair-produced mesons in representations larger than the $\mathbf{5}$ that hop down to lower representations may lead to interesting signals from emission of multiple electroweak bosons, and measuring the rate for such processes would help reconstruct $f_\pi$.

We focused on the light quark limit with $4\pi f_\pi\gtrsim2\,\text{TeV}$ because of the simplicity of the meson EFT. A smaller $f_\pi$ would result in complications due to having a tower of lighter excited meson states. A natural direction for future studies would be 
to include consideration of the vector mesons in our model, along the lines of the discussion in Refs.~\cite{Kribs:2018oad} and \cite{Kribs:2018ilo} for a similar dark sector. Our dark $\rhoDark$ mesons can mix with the SM $W$ and $Z$, leading to resonance signals that may be competitive with those discussed in this work when $f_\pi$ is sufficiently small. The flavor-singlet $\etaDark^\prime$ may also play an important role as an additional anomaly-induced resonance.

Another small modification to our regime is the heavy quark limit (studied in Ref.~\cite{Asadi:2024tpu} in the context of dark baryons). Here, producing dark sector states at a collider would probe the theory at a scale beyond the meson EFT's regime of validity. Instead, one would produce quirks \cite{Kang:2008ea}, 
where a quark/anti-quark pair is joined by an unbroken flux tube and non-perturbatively radiates dark glueballs, photons, \emph{etc.}~before annihilating. Such a regime would have distinct signals from those presented here that are also more difficult to model. An advantage of heavy quarks is the ability to compute the hadron mass spectrum in perturbation theory, thereby avoiding ambiguities such as the $\piDark_3$-$\etaDark_5$ mass splitting discussed above. One can also exploit the squeezeout mechanism in the heavy quark limit for determining the dark matter relic abundance of the dark baryons \cite{Asadi:2021pwo,Asadi:2021yml,Asadi:2022vkc}. (In either the heavy- or light-quark limit, a matter/anti-matter asymmetry in the dark sector may be necessary for the dark baryons to have the relic abundance of dark matter \cite{Mitridate:2017oky,Asadi:2021pwo,Asadi:2021yml,Asadi:2022vkc}.) The light and heavy quark limits are therefore qualitatively different, and each merits further investigation.

The work presented here shows an example of how dark matter models with $\mathcal{H}$-parity can be (counter-intuitively) more easily discovered at colliders than other experiments searching for dark sectors with electroweak interactions. The 5-plet anomaly also serves as an example of how the rich spectra of confining dark sectors can enable a variety of counter-intuitive search strategies for physics beyond the Standard Model. 

\section*{Acknowledgments}

We thank Tom Bouley, Spencer Chang, Tim Cohen, Adam Martin, and Nick Rodd for helpful discussions. The work of PA, AB, and GK is supported in part by the U.S. Department of Energy under grant number DE-SC0011640. 
EB is supported, in part, by the US National Science Foundation under Grant PHY-2210177.
MC is supported in part by Perimeter Institute for Theoretical Physics. Research at Perimeter Institute is supported by the Government of Canada through the Department of Innovation, Science and Economic Development Canada and by the Province of Ontario through
the Ministry of Research, Innovation and Science. 
SH is supported by the NSF grant PHY-2309456 and in part by the US-Israeli BSF grant 2016153.
This research was supported in part by grant NSF PHY-2309135 to the Kavli Institute for Theoretical Physics (KITP). 
PA is grateful to the theory group at UC Santa Cruz for their kind hospitality and stimulating environment.

\clearpage

\appendix

\section{Review of Standard Model Meson Anomalies}
\label{sec:SManomalies}

We summarize here the basics of the chiral anomalies that appear in QCD, which serve as an instructive comparison to the dark sector. 
They arise due to non-conservation of the chiral flavor currents $j_\mu^{5}$, such as the current corresponding to the axial U(1) mentioned in \cref{sec:chiral}. Consider the divergence of $j_\mu^{5a}$ due to SM gluons, with $a$ indexing the generators that are broken during confinement:
\begin{equation}
    \label{eq:QCDanomaly}
    \partial^\mu j^{5a}_{\mu} \propto G^b_{\text{SM}} \widetilde{G}^c_{\text{SM}} \text{Tr}\left[ T^a \,t^b\, t^c \right],
\end{equation}
\noindent where $G^a_{\text{SM}}$ is the QCD field strength, and $t^a$ is a generator of SU(3)$_{\text{color}}$. The flavor generator $T^a$ has flavor indices, while $t^a$ has color indices, so the trace factorizes as 
\begin{equation}
    \label{eq:traceFactorize}
    \text{Tr}\left[ T^a \,t^b\, t^c \right] = \text{Tr}\left[ T^a\right] \text{Tr}\left[t^b\, t^c \right]=0,
\end{equation}
\noindent since the flavor generator is traceless. However, if we consider the flavor-singlet $\eta^\prime_{\text{SM}}$, the ``flavor generator" is the identity in flavor space, which is not traceless, and the divergence becomes 
\begin{equation}
    \label{eq:QCDetaprime}
    \partial^\mu j^{5}_{\mu}  \propto G^a_{\text{SM}} \widetilde{G}^b_{\text{SM}} \text{Tr}\left[t^a\, t^b \right] = \frac{1}{2} G^a_{\text{SM}} \widetilde{G}^a_{\text{SM}}.
\end{equation}
\noindent The flavor-singlet axial U(1) is therefore not a good symmetry of the UV theory, so the $\eta^\prime_{\text{SM}}$ is not a pNGB.

The SM pNGBs do not have an anomaly with QCD, but the neutral pion famously decays via an anomaly with QED. The effective operator governing this decay is
\begin{equation}
    \label{eq:QEDanomaly}
    \mathcal{L}^{\text{QED}}_{\text{anomaly}} \propto \pi_{\text{SM}}^0 F \widetilde{F}\, \text{Tr}\left[ T^{\,0} \,Q^2 \right],
\end{equation}
\noindent where $F$ is the QED field strength, and the generators are 
\begin{align}
    T^{\,0} &= \frac{1}{2} 
    \begin{pmatrix}
        1 & & \\ & -1 & \\ & & 0
    \end{pmatrix},
    & 
    Q &= \begin{pmatrix}
        2/3 & & \\ & -1/3 & \\ & & -1/3
    \end{pmatrix},
\end{align}
\noindent which both have SU(3) flavor indices. The product $T^{\,0}\,Q^2$ is not traceless, so this anomaly is non-vanishing.

We emphasize that the above anomalies are non-vanishing even though QED and QCD are vector-like theories. The flavor currents appearing in triangle diagrams of the form 

\begin{center}
\includegraphics[width=0.3\textwidth]{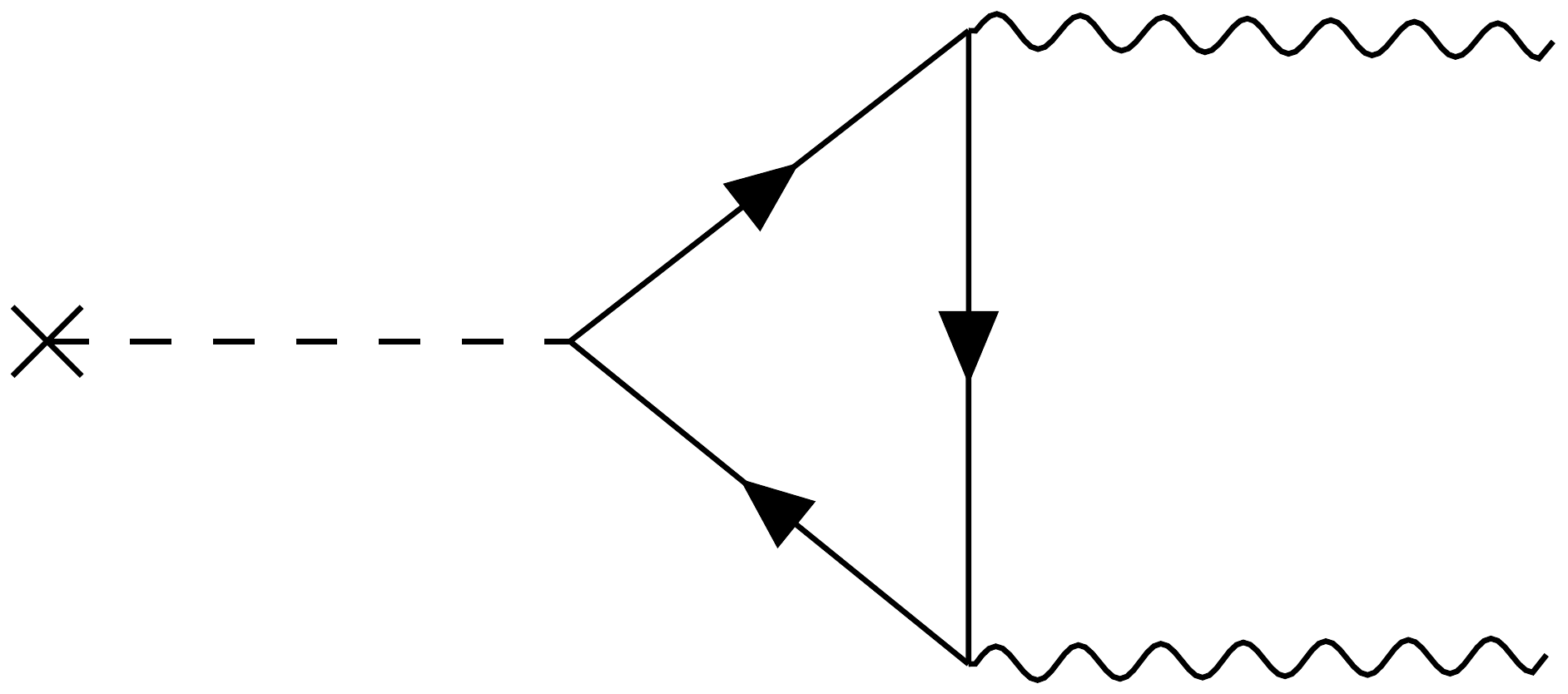}
\end{center}

\noindent are global chiral currents, so these global-gauge-gauge anomalies do not contradict the required absence of pure gauge anomalies. Each of these points is useful to keep in mind 
when discussing the analogous anomalies in the dark sector.
\section{Uniqueness of the 5-plet Anomaly}
\label{sec:uniqueness}

There is a simple argument from gauge invariance that the $\etaDark_5$ is unique among the pNGBs in that it has an anomaly with SU(2)$_L$. We have an effective operator in \cref{eq:Lanomaly} of the form $\piDark W\widetilde{W}$ or $\etaDark \, W\widetilde{W}$. The $W$ operators are in the $\mathbf{3}$ of SU(2)$_L$, and the $\piDark$ or $\etaDark$ is in some representation $\mathbf{d}$. The effective operator can only be a gauge singlet if the product
\begin{equation}
    \mathbf{3} \otimes \mathbf{3} \otimes \mathbf{d} = (\mathbf{1} \oplus \mathbf{3} \oplus \mathbf{5}) \otimes \mathbf{d}
\end{equation}
\noindent contains a $\mathbf{1}$. This only occurs if $\mathbf{d}$ is $\mathbf{1}$, $\mathbf{3}$, or $\mathbf{5}$, so only the $\etaDark^\prime$, $\piDark_3$, and $\etaDark_5$ could have the anomaly. In light of the discussion in \cref{sec:SManomalies}, it is unsurprising that the $\etaDark^\prime$ has an anomaly with SU(2)$_L$ in addition to SU($N_c$), but it is not a pNGB. This leaves only the $\piDark_3$ and $\etaDark_5$, but the anomaly with the $\piDark_3$ vanishes for other reasons explained below.\footnote{
Outside of the context of our model, one could  consider a minimal dark matter candidate that is the neutral component of a pseudoscalar SU(2)$_L$ 5-plet $\phi_5$. Using the UV-agnostic framework of Ref.~\cite{Cirelli:2005uq}, the $\phi_5 W \widetilde{W}$ effective operator (analogous to the anomaly operator in this work) would destabilize this dark matter candidate, and additional model building would be required for the UV completion to forbid this decay channel. This observation builds on the result of Ref.~\cite{Cirelli:2005uq} that a \textit{scalar} 5-plet minimal dark matter candidate can also be destabilized at dimension-5.
}

If the term in \cref{eq:Lanomaly} with the $\piDark_3$ were non-vanishing, then it would clearly violate $G$-parity. However, we show here that this term vanishes regardless of $G$-violation in the UV (which further implies that the anomaly preserves $G$). Consider how one would determine the flavor generator $T^a$ in \cref{eq:Lanomaly} corresponding to a particular meson species in a particular multiplet. The flavor generator encodes the meson state as a linear combination of quark/anti-quark flavor states, which should be an eigenstate of the SU(2)$_L$ quadratic Casimir operator $J^2$ and the electric charge operator $Q$. Transformations of the meson state under SU(2)$_L$ can be written as transformations of $T^a$ using 
\begin{equation}
    T^a\stackrel{\vphantom{\int}J^i}{\to} [J^i,T^a],
\end{equation}
\noindent where $J^i$ is a generator of the $N_f$-dimensional representation of SU(2)$_L$. Then, the charge and Casimir operators act as
\begin{align}
    T^a &\stackrel{\vphantom{\int}Q}{\to} [J_z,T^a], & T^a &\stackrel{\vphantom{\int}J^2}{\to} [J^i,[J^i,T^a]]\,.
\end{align}
\noindent Therefore, the meson with charge $Q$ in the $2J+1$-dimensional representation of SU(2)$_L$ corresponds to the flavor generator with 
\begin{align}
    \label{eq:eigenGenerators}
    T^a &\stackrel{\vphantom{\int}Q}{\to} Q\,T^a, & T^a &\stackrel{\vphantom{\int}J^2}{\to} J(J+1)\,T^a.
\end{align}
\noindent We provide a recipe to explicitly construct these generators in \cref{sec:generators}. Suppose a meson's flavor generator is itself a linear combination of the SU(2)$_L$ generators. Then, its transformation under the Casimir operator is found using
\begin{equation}
    J^a \stackrel{\vphantom{\int}J^2}{\to} [J^i,[J^i,J^a]] = \varepsilon^{aij}\varepsilon^{ijk}J^k = 2J^a,
\end{equation}
\noindent \textit{i.e.}~the meson is in the $J=1$ (3-plet) representation. Thus, the flavor generators corresponding to the $\piDark_3$ multiplet are (unsurprisingly) linear combinations of SU(2)$_L$ generators. Then, the 
operator in \cref{eq:Lanomaly} for the $\piDark_3$ contains
\begin{equation}
    \label{eq:tripletTrace}
    \varepsilon^{\mu\nu\alpha\beta} W^i_{\mu\nu} W^j_{\alpha\beta} \text{Tr} \left[ J^a J^i J^j  \right].
\end{equation}
\noindent Unlike the SM flavor and color generators in \cref{eq:traceFactorize}, the generators in this trace all have the same flavor indices (analogously to \cref{eq:QEDanomaly}), so the trace itself neither factorizes nor vanishes. However, the trace is anti-symmetric 
in the $i$ and $j$ indices,\footnote{Equivalently, the totally symmetric invariant tensor $d^{abc}$ of SU(2) vanishes.} while $\varepsilon^{\mu\nu\alpha\beta} W^i_{\mu\nu} W^j_{\alpha\beta}$ is symmetric in $i$ and $j$, so the operator vanishes. Therefore, the $\piDark_3$ does not have an anomaly with SU(2)$_L$. The flavor generators for meson multiplets besides the $\piDark_3$ are \emph{not} linear combinations of SU(2)$_L$ generators, so the above argument does not forbid the $\etaDark_5$ from having the anomaly. 

In this way, we have shown that the $\etaDark_5$ is the unique pNGB with 
a global-SU(2)$_L$-SU(2)$_L$ anomaly. One can also see this by finding the generators for the $\piDark_7,\etaDark_{\,9}$, \emph{etc.}~and observing that the trace in \cref{eq:Lanomaly} for these species vanishes identically, unlike the trace in \cref{eq:tripletTrace}. 

\section{Obtaining Flavor Generators}
\label{sec:generators}

Here, we present a method to explicitly construct the flavor generators corresponding to meson species in specific SU(2)$_L$ multiplets with specific electric charges. One starts with a basis for the diagonal generators $\{D^a\}$ of SU($N_f$). For example,
\begin{align}
    D^1 &= \frac{1}{2}\, \text{diag}(1,-1,0,\dots,0), \\
    D^2 &= \frac{1}{2\sqrt{3}}\, \text{diag}(1,1,-2,0,\dots,0), \\
    &\dots, \\
    D^{N_f-1} &= \frac{1}{\sqrt{2N_f(N_f-1)}}\,\text{diag}(1,\dots,1,1-N_f)\,.
\end{align}
\noindent Then, one can define the matrix
\begin{equation}
    M_{ab} = 2\,\text{Tr}\left( D^a [J^i,[J^i,D^b]] \right),
\end{equation}
\noindent where $J^i$ is a generator of the $N_f$-dimensional representation of SU(2)$_L$. As discussed in \cref{sec:uniqueness}, the matrix $M$ encodes the action of the quadratic Casimir operator of SU(2)$_L$ on the generators. The eigenvalues of $M$ are $J(J+1)$ with $J\in\{1,2,\dots,N_f-1\}$, and the eigenvectors provide the linear combinations of the $D^a$ generators corresponding to the neutral mesons in each of the SU(2)$_L$ representations $\mathbf{3},\mathbf{5},\dots,\mathbf{2N_f-1}$. They must be the neutral mesons because the generators are diagonal and therefore commute with the third SU(2)$_L$ generator $J_z$. Now that one has the generators for the neutral mesons in each representation, one can construct the remaining flavor generators using the SU(2)$_L$ ladder operators 
\begin{align}
    J^\pm &= J_x\pm i J_y, \\
    \langle m_1 | J^\pm | m_2\rangle &= \delta_{m_1,m_2\pm1}\sqrt{\frac{N_f^2-1}{4}-m_2(m_2\pm1)}\,,
\end{align}
\noindent where $J_x$ and $J_y$ are the first and second SU(2)$_L$ generators. If $T_J^{\,Q}$ is the flavor generator corresponding to the meson with charge $Q$ in the $2J+1$-dimensional representation of SU(2)$_L$, then one can raise or lower it to obtain
\begin{equation}
    T_J^{\,Q\pm1} = \frac{1}{\sqrt{J(J+1)-Q(Q\pm1)}} [J^\pm,T_J^{\,Q}]\,.\label{eq:raiseLower}
\end{equation}
\noindent Each multiplet contains a neutral meson, so the iterative application of the ladder operators to the diagonal generators allows one to find the flavor generators for all meson species.

As an explicit example, consider the generators for the meson 5-plet $\etaDark_5$ when $N_f=3$. A basis for the diagonal generators of SU(3) is 
\begin{align}
    D^1 &= \frac{1}{2}
    \begin{pmatrix}
        1 & & \\
        & -1 & \\
        & & 0
    \end{pmatrix}, &
    D^2 &= \frac{1}{2\sqrt{3}}
    \begin{pmatrix}
        1 & & \\
        & 1 & \\
        & & -1
    \end{pmatrix},
\end{align}
\noindent with which one finds
\begin{equation}
    M = 
    \begin{pmatrix}
        5 & -\sqrt{3} \\
        -\sqrt{3} & 3
    \end{pmatrix}.
\end{equation}
\noindent $M$ has eigenvalues 2 and 6, as we expect from having 3-plet and 5-plet mesons. The eigenvectors of $M$ show that the generator corresponding to the neutral 5-plet meson is 
\begin{equation}
    T_2^{\,0} = \frac{\sqrt{3}}{2}D^1 - \frac{1}{2}D^2 = \frac{1}{2\sqrt{3}}
    \begin{pmatrix}
        1 & & \\
        & -2 & \\ 
        & & 1
    \end{pmatrix}.
\end{equation}
\noindent Then, using the ladder operators 
\begin{align}
J^+ &=\sqrt{2}
\begin{pmatrix}
    0 & 1 & 0 \\
    0 & 0 & 1 \\
    0 & 0 & 0
\end{pmatrix},
&
J^- &=\sqrt{2}
\begin{pmatrix}
    0 & 0 & 0 \\
    1 & 0 & 0 \\
    0 & 1 & 0
\end{pmatrix}
\end{align}
\noindent in combination with \cref{eq:raiseLower}, one can construct the remaining 5-plet flavor generators
\begin{align}
    T_2^{++} &= \frac{1}{\sqrt{2}}
    \begin{pmatrix}
        0 & 0 & 1 \\
        0 & 0 & 0 \\
        0 & 0 & 0
    \end{pmatrix},
    &
    T_2^{+} &= \frac{1}{2}
    \begin{pmatrix}
        0 & -1 & 0 \\
        0 & 0 & 1 \\
        0 & 0 & 0
    \end{pmatrix}, \\
    T_2^{--} &= \frac{1}{\sqrt{2}}
    \begin{pmatrix}
        0 & 0 & 0 \\
        0 & 0 & 0 \\
        1 & 0 & 0
    \end{pmatrix},
    &
    T_2^{-} &= \frac{1}{2}
    \begin{pmatrix}
        0 & 0 & 0 \\
        1 & 0 & 0 \\
        0 & -1 & 0
    \end{pmatrix}.
\end{align}
\noindent This calculation helps establish the relative factors between different terms in the anomaly operator in \cref{eq:5pletAnomaly}, and the overall factor $c$ is computed in \cref{sec:c}.

\section{Computing the 5-plet Anomaly}
\label{sec:c}

The coefficient $c$ in \cref{eq:5pletAnomaly} encodes the $N_f$ dependence of the 5-plet anomaly. Here, we compute $c$ for an arbitrary $N_f$ using general properties of the 5-plet generators. To start with a helpful analogy, recall from \cref{sec:uniqueness} that the generators corresponding to the 3-plet mesons are linear combinations of the SU(2)$_L$ generators $J^i$ in the $N_f$-dimensional representation. This is expected, as $J^i$ transforms as a $\mathbf{3}$, so the generators of the $\mathbf{3}$ do so as well. 

We can make a similar argument for the 5-plet by noticing that the $\mathbf{5}$ of SU(2) is the symmetric traceless combination of two $\mathbf{3}$'s. Therefore, the object
\begin{equation}
    S^{ij} = \{J^i,J^j\}-\frac{1}{N_f}\,\text{Tr}(\{J^i,J^j\})\mathds{1},
\end{equation}
\noindent where $\mathds{1}$ is the identity in flavor space, is a tensor that transforms as a $\mathbf{5}$. This means that the generators for the 5-plet mesons are linear combinations of the components of $S^{ij}$. To fully convince oneself of this, one can explicitly show 
\begin{equation}
    [J^k,[J^k,S^{ij}]] = 6S^{ij},
\end{equation}
\noindent as required of the 5-plet generators (see \cref{eq:eigenGenerators}), using the identities 
\begin{gather}
    [J^k,[J^k,\{J^i,J^j\}]] = 6\{J^i,J^j\}-4\delta^{ij}J^2 = 6\{J^i,J^j\}-(N_f^2-1)\delta^{ij}\, \mathds{1},\\
    \text{and} \notag \\ 
    \frac{1}{N_f}\,\text{Tr}(\{J^i,J^j\}) = \frac{1}{6}(N_f^2-1)\delta^{ij}\,.
\end{gather}
\noindent Then, the flavor generators corresponding to each 5-plet species are proportional to the components of $S^{ij}$ in the spherical basis. In particular, the diagonal generator (corresponding to the neutral component of the 5-plet) is \cite{Varshalovich:1988ifq,Sakurai:2011zz}
\begin{equation}
\begin{aligned}
    T_2^0&\propto \frac{1}{\sqrt{6}}(2S^{33}-S^{11}-S^{22}) \\ 
    &\propto 3J_z^2 - J^2 = 3J_z^2 - \frac{N_f^2-1}{4}\, \mathds{1}\,.
\end{aligned}
\end{equation}
\noindent To enforce the normalization Tr$((T_2^0)^2)=1/2$, we have 
\begin{equation}
    T_2^0 = \sqrt{10\frac{(N_f-3)!}{(N_f+2)!}} \left(3J_z^2 - 
    \frac{N_f^2-1}{4}\, \mathds{1}
    \right),
\end{equation}
\noindent which follows from
\begin{equation}
    J_z=\text{diag}((N_f-1)/2,(N_f-1)/2-1,\dots,-(N_f-1)/2)\,.
\end{equation}
\noindent The general form for $T_2^0$ will help us compute $c$.

The anomaly operator in \cref{eq:Lanomaly} contains traces of the form 
\begin{equation}
    \text{Tr}(T^aJ^iJ^j)\,.
\end{equation}
\noindent The $a$ index picks out a particular meson species, and the $i$ and $j$ indices pick out particular components of the SU(2)$_L$ field strength $W_{\mu\nu}^i$. If we focus on the term with $i=j=3$, we have 
\begin{equation}
    \label{eq:anomaly33}
    \mathcal{L}_{\text{anomaly}}\supset \frac{g_W^2}{16\pi^2}\frac{N_c}{f_\pi} \varepsilon^{\mu\nu\alpha\beta}W_{\mu\nu}^3W_{\alpha\beta}^3\,\etaDark_5^0\,\text{Tr}(T_2^0J_z^2)\,.
\end{equation}
\noindent The relative factors between the traces over different combinations of generators are fixed by gauge invariance, so it is sufficient to compute only one of them for an arbitrary $N_f$, and $c$ is an overall factor in the anomaly operator. Thus, we can match \cref{eq:anomaly33} onto \cref{eq:5pletAnomaly} to conclude 
\begin{equation}
\begin{aligned}
    c &= \frac{1}{2} \text{Tr}(T_2^0J_z^2) \\
    &= \frac{1}{12\sqrt{10}}\sqrt{\frac{(N_f+2)!}{(N_f-3)!}}\,.
\end{aligned}
\end{equation}
\noindent This tells us how the strength of the anomaly interaction varies with $N_f$.

\bibliographystyle{JHEP}
\bibliography{HCollider}

\end{document}